\documentclass[11pt,a4paper]{article}
\usepackage{cite}

\usepackage{amsmath, amsthm,mathtools,setspace, hyperref, url,amssymb,upgreek,textgreek,slashed}

 \usepackage[mathscr]{eucal}
\usepackage{ifpdf}
\ifpdf
  \usepackage[pdftex]{graphicx}
  \usepackage{epstopdf}
\else
  \usepackage[dvips]{graphicx}
\fi
\parskip=\baselineskip

\begin{document}

\begin{titlepage}
\begin{flushright}

\end{flushright}

\vskip 1.5in
\begin{center}
{\bf\Large{A Note On Complex Spacetime Metrics}}

\vskip
0.5cm  { Edward Witten} \vskip 0.05in {\small{ \textit{Institute for Advanced Study}\vskip -.4cm
{\textit{Einstein Drive, Princeton, NJ 08540 USA}}}
}
\end{center}
\vskip 0.5in
\baselineskip 16pt
\begin{abstract}  For various reasons, it seems necessary to include complex saddle points in the ``Euclidean'' path integral of General Relativity.
But some sort of restriction on the allowed complex saddle points is needed to avoid various unphysical examples.   In this article, a 
speculative proposal is made concerning a possible restriction on the allowed saddle points in the gravitational path integral.   The proposal
is motivated by recent work of Kontsevich and Segal on complex metrics in quantum field theory, and earlier work of Louko and Sorkin on topology
change from a real time point of view.
   \end{abstract}
\date{October, 2021}
\end{titlepage}
\def\SO{{\mathrm{SO}}}
\def\G{{\text{\sf G}}}
\def\frak{\mathfrak}
\def\la{\langle}
\def\ra{\rangle}
\def\Spinc{{\mathrm{Spin}}_c}
\def\g{{\mathfrak g}}
\def\cl{{\mathrm{cl}}}
\def\m{{\sf m}}
\def\veps{\varepsilon}
\def\Re{{\mathrm{Re}}}
\def\Im{{\mathrm{Im}}}
\def\SU{{\mathrm{SU}}}
\def\SL{{\mathrm{SL}}}

\def\M{{\mathcal M}}
\def\d{{\mathrm d}}
\def\g{{\mathfrak g}}
\def\su{{\mathfrak {su}}}
\def\CS{{\mathrm{CS}}}
\def\Z{{\Bbb Z}}
\def\cB{{\mathcal B}}
\def\zZ{{\mathcal Z}}
\def\DD{{\mathscr D}}
\def\R{{\Bbb R}}
\def\sF{{\sf F}}
\def\cS{{\mathcal S}}
\def\sB{{\sf B}}
\def\sA{{\sf A}}
\def\sD{{\mathcal D}}
\def\dD{{\mathrm D}}
\def\F{{ \mathscr F}}
\def\cF{{\mathcal F}}
\def\J{{\mathcal J}}
\def\Bbb{\mathbb}
\def\Tr{{\rm Tr}}
\def\ad{{\mathrm{ad}}}
\def\j{{\sf j}}
\def\16{{\bf 16}}
\def\1{{(1)}}
\def\bCP{{\Bbb{CP}}}
\def\2{{(2)}}
\def\3{{\bf 3}}
\def\4{{\bf 4}}
\def\free{{\mathrm{free}}}
\def\sg{{\mathrm g}}
\def\J{{\mathcal J}}
\def\i{{\mathrm i}}
\def\h{\widehat}
\def\b{\overline}
\def\u{u}
\def\D{D}
\def\Rf{{\eurm{R}}}
\def\sp{{\sigma}}
\def\E{{\mathcal E}}
\def\O{{\mathcal O}}
\def\OO{{\mathrm O}}
\def\PF{{\mathit{P}\negthinspace\mathit{F}}}
\def\tr{{\mathrm{tr}}}
\def\be{\begin{equation}}
\def\ee{\end{equation}}
 \def\Sp{{\mathrm{Sp}}}
  \def\PSp{{\mathrm{PSp}}}
 \def\Spin{{\mathrm{Spin}}}
 \def\SL{{\mathrm{SL}}}
 \def\SU{{\mathrm{SU}}}
 \def\SO{{\mathrm{SO}}}
 \def\PGL{{\mathrm{PGL}}}
 \def\ll{\langle\langle}
 \def\frL{{{\mathfrak L}}}
 \def\RR{{\mathcal R}}
\def\rr{\rangle\rangle}
\def\la{\langle}
\def\CP{{C\negthinspace P}}
\def\sCP{{\sf{CP}}}
\def\I{{\mathcal I}}
\def\ra{\rangle}
\def\T{{\mathcal T}}
\def\V{{\mathcal V}}
\def\bar{\overline}
\def\spinc{{\mathrm{spin}_c}}
\def\dim{{\mathrm{dim}}}
\def\v{v}
\def\Pic{{\mathrm{Pic}}}

\def\RP{{\Bbb{RP}}}
\def\C{{\mathbb C}}
\def\tilde{\widetilde}
\def\t{\widetilde}
\def\R{{\Bbb{R}}}
\def\N{{\mathcal N}}
\def\B{{\mathcal B}}
\def\H{{\mathcal H}}
\def\hat{\widehat}
\def\Pf{{\mathrm{Pf}}}
\def\bM{{\overline\M}}
\def\PSL{{\mathrm{PSL}}}
\def\PSU{{\mathrm{PSU}}}
\def\Im{{\mathrm{Im}}}
\def\Gr{{\mathrm{Gr}}}
\def\sign{{\mathrm{sign}}}

\font\tencmmib=cmmib10 \skewchar\tencmmib='177
\font\sevencmmib=cmmib7 \skewchar\sevencmmib='177
\font\fivecmmib=cmmib5 \skewchar\fivecmmib='177
\newfam\cmmibfam
\textfont\cmmibfam=\tencmmib \scriptfont\cmmibfam=\sevencmmib
\scriptscriptfont\cmmibfam=\fivecmmib
\def\cmmib#1{{\fam\cmmibfam\relax#1}}
\numberwithin{equation}{section}
\def\lmark{{\mathrm L}}
\def\neg{\negthinspace}

\def\C{{\Bbb C}}
\def\K{{\sf K}}
\def\cc{{\mathrm{cc}}}
\def\bB{{\bar {\mathscr B}}}
\def\HH{{\mathbb H}}
\def\P{{\mathcal P}}
\def\Q{{\mathcal Q}}
\def\NS{{\sf{NS}}}
\def\Ra{{\sf{R}}}
\def\sV{{\sf V}}
\def\CP{{\mathrm{CP}}}
\def\Hom{{\mathrm{Hom}}}
\def\Z{{\Bbb Z}}
\def\op{{\mathrm{op}}}
\def\bop{{\overline{\mathrm{op}}}}
\def\bA{\bar{\mathscr A}}
\def\A{{\mathscr A}}
\def\cA{{\mathcal A}}
\def\B{{\mathscr B}}
\def\S{{\sf S}}
\def\bar{\overline}
\def\sc{{\mathrm{sc}}}
\def\Max{{\mathrm{Max}}}
\def\CS{{\mathrm{CS}}}
\def\ga{\gamma}
\def\bg{\bar\ga}
\def\Arg{{\mathrm{Arg}}}
\def\W{{\mathcal W}}
\def\M{{\mathcal M}}
\def\bM{{\overline \M}}
\def\L{{\mathcal L}}
\def\sM{{\sf M}}
\def\gst{\mathrm{g}_{\mathrm{st}}}
\def\gstt{\widetilde{\mathrm{g}}_{\mathrm{st}}}
\def\hbbar{\pmb{\hbar}}
\def\G{{\mathcal G}}
\def\Sym{{\mathrm{Sym}}}
\def\U{{\mathcal U}}
\def\Bun{{\mathcal M}(G,C)}
\def\be{\begin{equation}}
\def\ee{\end{equation}}
\def\Diff{{\mathrm{Diff}}}
\def\diff{{\mathrm{diff}}}

\tableofcontents

\section{Introduction}\label{intro}

In their original paper interpreting black hole entropy in terms of the gravitational action,  Gibbons and Hawking \cite{GH}, after analyzing the thermodynamics of a Schwarzschild black hole,   went on
to consider   black holes with angular momentum.   They pointed out that the Kerr metric, assuming that the angular momentum is real, 
becomes complex-valued  when continued to imaginary time.  While complex-valued, this metric is everywhere nondegenerate, like all complex
metrics that will be considered in this article.   Gibbons and Hawking   showed that one can recover the expected results for the thermodynamics 
of the Kerr solution by assuming that this complex saddle point dominates the appropriate path integral.

Somewhat later, Gibbons, Hawking, and Perry \cite{GHP}, observing that the action of Euclidean quantum gravity is not positive-definite,  argued that
therefore the path integral of ``Euclidean'' quantum gravity should really be understood as a sort of infinite-dimensional version of a complex
contour integral, with the integration running over a suitable family of nondegenerate complex metrics on spacetime.     

Since then,  other reasons have been put forward to consider complex spacetime metrics and complex saddle points.   For example,
it  has been argued by Halliwell and Hartle \cite{HalHar}
that to get sensible answers for the behavior of large, semiclassical spacetimes from the Hartle-Hawking no-boundary proposal for the wavefunction of the universe \cite{HH},
one must consider complex solutions of Einstein's equations as saddle points.   It has also been argued by Louko and Sorkin \cite{SL} that to get sensible answers
for real time topology-changing processes, one must consider complex spacetime metrics that correspond roughly to tunneling trajectories.  Various additional arguments have been given, some of which will be discussed later.

On the other hand, in considering complex saddle points of Einstein's equations, one is potentially opening Pandora's box.   Many
such saddle points,  if included in a functional integral, will give results that are not physically sensible.    This point was made in \cite{SL},
and we will illustrate it further in section \ref{examples} with some additional examples.
The examples of section \ref{examples} are all constructed by taking real submanifolds of simple complex manifolds, so perhaps we should point
out that this method of constructing examples is useful but in a sense atypical.   There typically is no canonical way to complexify a real manifold $M$ or to analytically
continue a complex metric $g$, and generically a complex metric on $M$ is not related in any useful way to a complexification of $M$.

If some complex metrics are ``bad,'' which ones are ``good''?
Consider a semiclassical theory of gravity coupled to  matter,  and assume that the matter is described by
ordinary quantum fields.   In that context, an important necessary condition  for complex metrics, discussed for example in
\cite{HalHar} and \cite{SL},  is that
the complex spacetime should be one  in which the quantum field theory of the matter system can be defined.  If the matter system is sufficiently generic,
 the complex spacetime should be one in which more or less any quantum field theory could be defined.

  Recently, with a different motivation,
Kontsevich and Segal \cite{KS} have made a proposal
for what is the class of complex geometries  in which a generic quantum field theory can be consistently coupled.    
Their proposal was not directly motivated by quantum gravity;
their basic goal was to explore the extent to which 
 traditional axiom sets of quantum field theory can be  replaced by the assumption that a theory can be consistently
coupled to a certain class of complex metrics.    However, it is interesting, though  speculative,
to consider their class of ``allowable'' complex metrics in the context of quantum gravity.  

 In section \ref{allowable} of this paper, we describe the class of metrics considered in \cite{KS} and show that the problematical examples of section \ref{examples}
 are not allowable in their sense.   Then in section \ref{useful}, we consider some of the cases in which apparently useful statements about
 quantum gravity have been made using complex solutions of Einstein's equations, and show that the metrics considered are allowable.  
 This gives at least some support for the idea of restricting to ``allowable'' metrics as saddle points.
   
In section \ref{cycle}, we discuss from this point of view the integration cycle of the gravitational path integral.    As originally discussed by Gibbons, Hawking, and Perry
\cite{GHP}, the action of Euclidean quantum gravity is not bounded below.    They proposed that the gravitational path integral should be understood as an integral
over a real cycle in the space of complex metrics.   In the context of perturbation theory around a given classical solution, their proposal is satisfactory, and,
if the given solution is allowable, their construction stays in the class of allowable metrics.   A satisfactory
 extension of their proposal beyond perturbation theory is not apparent.

Finally, in Appendix \ref{gb}, we analyze and generalize a question from \cite{SL}.  The original question involved the Euler characteristic of a two-manifold $M$ and
the conditions under which it can be computed as $\int_M\d^2 x \sqrt{\det g}R/4\pi$, where $R$ is the Ricci scalar of a complex metric $g$.

It is a pleasure to submit this article to a volume in honor of the 70th birthday of Frank Wilczek, who was my colleague  at Princeton in the 1970's,
and later on the IAS faculty.  Frank has made many important contributions to different areas of physics, never shying away from bold speculation.  So
I hope it is not too inappropriate to submit to this volume a rather speculative article, which is largely motivated by a limited number of examples that are discussed
in section \ref{useful}.\footnote{An earlier version of part of this material was presented at a conference
in honor of the 70th birthday of another distinguished physicist, Thibault Damour, who also was a colleague of Frank and me at Princeton in the mid-1970's.}

\section{Some Examples of Complex Solutions of Einstein's Equations}\label{examples}

The usual flat metric on $\R^D$ can be written
\be\label{flatmet}\d s^2=\d r^2+r^2\d\Omega^2, \ee
where $r$ is a non-negative real variable and $\d\Omega^2$ is the usual round metric on a sphere  $\S^{D-1}$.
One simple way to generalize this to a complex invertible metric,\footnote{In $D=2$, this example was briefly described in footnote 5 of \cite{SL}.} 
is to leave alone the real sphere $\S^{D-1}$ but relax the condition for $r$ to be real.
Instead, we specify that $r$ runs over a curve in the complex $r$-plane, for example a curve $r=r(u)$, where $u$ is a  real variable.   
The metric is then 
\be\label{inflat} \d s^2= r'(u)^2\d u^2+ r(u)^2 \d\Omega^2 \ee
This metric is nondegenerate as long as $r(u)$ and $r'(u)$ are both nonzero for all $u$.    
In that case, the metric is flat.  Indeed, if $r(u)$ is real, the metric (\ref{inflat})  just differs from the original flat metric (\ref{flatmet})
by a reparametrization, so it is certainly flat.   But when one verifies this flatness, one never has to use the fact that $r(u)$ is real, so the metric remains
flat even when $r(u)$ is complex-valued.

\begin{figure}
 \begin{center}
   \includegraphics[width=5in]{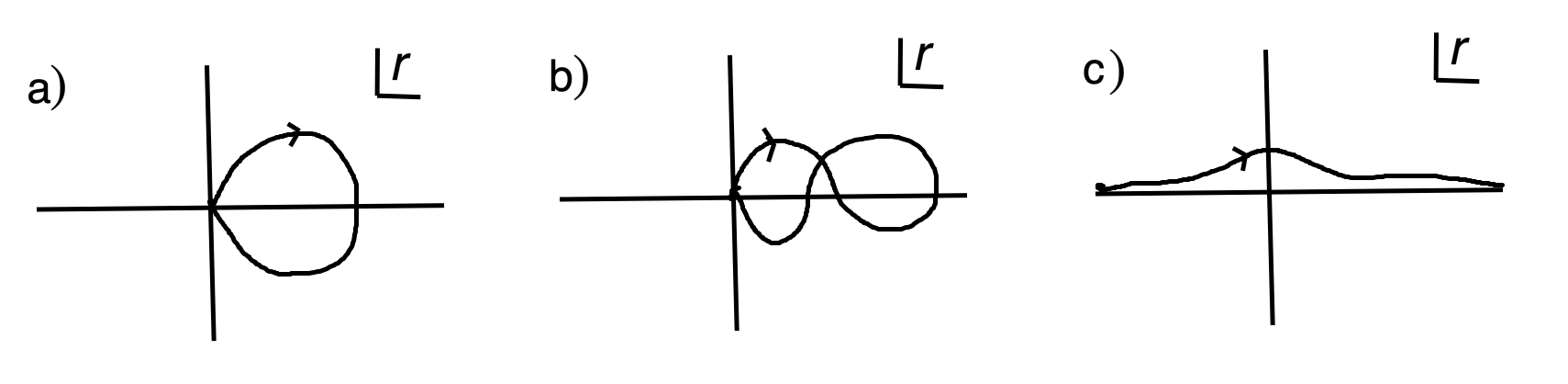}
 \end{center}
\caption{\small Some choices of curve $r(u)$ in the complex $r$-plane, leading to various complex flat metrics. (a) A loop that starts and ends at $r=0$,
leading to a complex  flat metric on $\S^D$.  (b) A homotopically inequivalent immersed loop that starts and ends at $r=0$, leading to
another complex flat metric on $\S^D$.  This example is equivalent to the one in (a) if equivalence is based on homology rather than homotopy.
 (c) A path from $r=-\infty$ to $r=+\infty$, avoiding the origin in the complex plane.   It leads to a flat ``wormhole''
 metric, as sketched in fig. \ref{Wormhole}(a). \label{FourChoices}}
\end{figure}

If we want a compact manifold without boundary, we should take $u$ to run over a compact interval $u_0\leq u\leq u_1$ and require
$r(u_0)=r(u_1)=0$, as in fig. \ref{FourChoices}(a).    Topologically this gives a sphere $\S^D$. 
   The complex metric on $\S^D$ that is obtained in this way  certainly satisfies the
Einstein equations with zero cosmological constant, since it is flat.

We get more options if we complexify the sphere $\S^{D-1}$.   A unit sphere defined by real variables $\vec x$ satisfying $\vec x^2=1$ can be complexified
by simply taking the components of $\vec x$ to be complex variables satisfying the same equation.   If we complexify $\S^{D-1}$ as well as the radial variable
$r$, we get a complex manifold $M_\C$ that comes with a holomorphically
varying, complex nondegenerate metric.  The metric of a real sphere can
be defined as $\d\Omega^2=\d \vec x\cdot \d \vec x$, with the constraint $\vec x\cdot \vec x=1$, and, after complexification, the same formula and constraint
define a nondegenerate and holomorphically varying complex
metric on the complexification of the sphere.  So a flat, holomorphic metric on $M_\C$ can be written as\footnote{As this formula suggests,
$M_\C$ is closely related to the usual complexification $\C^D$ of $\R^D$, which we could have used, albeit less interestingly, in this discussion.   It is more
convenient to proceed as in the text.} $\d r^2+r^2\d\Omega^2$.   By picking a curve $r(u)$ while
also keeping the angular variables real, we have described in the last paragraph an embedding of a real $D$-manifold $M\cong \S^D$ in $M_\C$.   Of course, we can consider more general embeddings, or even immersions,\footnote{An immersion is
a map that is locally an embedding.  In the examples described previously, the curve $r(u)$ might be immersed rather than embedded in the complex
$r$-plane; this will suffice to give a nondegenerate complex metric.} of $M$  in $M_\C$, with the angular variables no longer real, and this will give
more general complex metrics on the same $M$, though the condition that the metric should be everywhere nondegenerate puts a strong constraint
on the immersion of $M$ in $M_\C$.  

What equivalence relation should we place on complex metrics derived from different immersions of a given $M$ in the same $M_\C$?
  If has been proposed that two complex metrics on $M$ obtained
 by different immersions in the same $M_\C$ should be considered equivalent if they differ by ``complex diffeomorphisms.''  However, it appears difficult to define
 this notion precisely. 
  It seems that at a minimum, we should insist that two metrics on $M$ are equivalent if they come from homotopic immersions of $M$ in the same $M_\C$.   More optimistically, one might hope that two immersions of $M$ in $M_\C$ give equivalent metrics if they
 are merely homologous.    Going back to the case that the angular variables are real, in our examples, the first condition would
 say that two metrics are equivalent if they come from 
  immersed curves $r(u)$ and $\tilde r(u)$ that are homotopic
 (keeping endpoints fixed) and the second condition would say that for equivalence of the metrics, the curves just have to be
   homologous.  An equivalence relation based on homology is much
 stronger than one based on homotopy.   This is  illustrated in fig. \ref{FourChoices}(b); there are
 many classes of curves $r(u)$ associated to complex flat metrics on $\S^D$ that are homologous but not homotopic. (It may be that some of these become
 homotopic once one allows the angular variables to become complex.)

To get exotic complex flat metrics on $\R^D$, we can simply use the same construction, but now with $u$ ranging over the semi-infinite interval $[0,\infty)$.
To get $\R^D$ topologically, we require $r(0)=0$; to get a metric asymptotic to the standard Euclidean metric on $\R^D$, we require
$r(u)\sim u$ for $u\to\infty$.   If we take $r(u)$ to be identically equal to $u$ for $u>c$ (for some constant $c$), we get a metric that  coincides with the standard
Euclidean metric outside a bounded region.   Of course, the case $r(u)=u$ just gives back the standard Euclidean metric.  
We can get other families of complex flat metrics on $\R^D$ by choosing the curve $r(u)$ to be immersed, rather than embedded, in the complex $r$ plane.
Provided that  $r(u)$ approaches $u$ sufficiently rapidly at infinity to avoid a boundary term in the Einstein action, all of these flat metrics on $\R^D$
have vanishing action, like the standard one.

 \begin{figure}
 \begin{center}
   \includegraphics[width=2.5in]{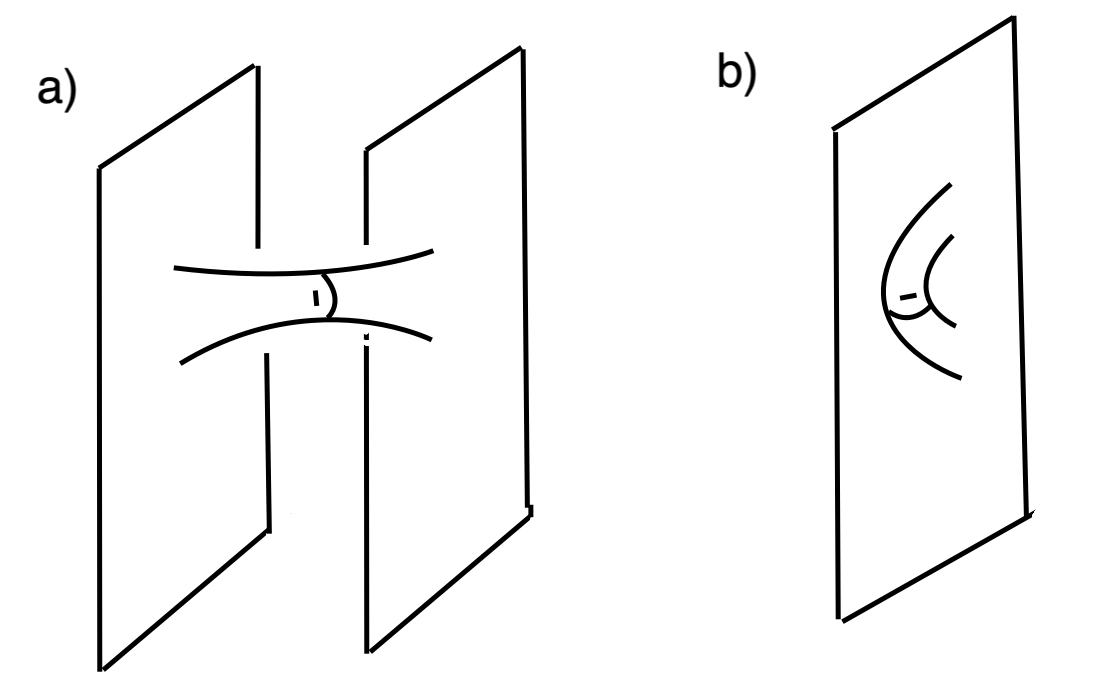}
 \end{center}
\caption{\small (a) Two copies of $\R^D$ have been connected via a ``wormhole.''  (b)  Two possibly distant regions of the same $\R^D$ are connected
by a wormhole.   \label{Wormhole}}
\end{figure} 

However, these examples are all equivalent if the appropriate notion of equivalence is based on homology.
More interesting are ``wormhole'' solutions, which we can get if
 $u$ runs over the whole real line.   We assume that $r(u)\to \pm\infty$ for $u\to \pm \infty$ and that $r(u)\not=0$ for all $u$.   A simple example is
 in fig. \ref{FourChoices}(c).     Such a construction gives 
  a connected spacetime (fig. \ref{Wormhole}(a)) with two
 ``ends'' each of which is asymptotic to a copy of $\R^D$; the ends  are connected through a wormhole. 
 So this construction gives complex wormhole solutions of  Einstein's equations with zero action.    For additional examples of the same
 type,  we could have $r(u)\to +\infty$ at both ends $u\to \pm\infty$; the curve $r(u)$ could be immersed rather
 than embedded, and in particular it could wrap any integer number of times around the point $r=0$.

Given a solution of Einstein's equations (possibly with matter fields)
in which a wormhole connects two different asymptotically flat
regions of spacetime, typically one can cut and paste to get an approximate solution in which a similar wormhole 
connects two distant regions of a spacetime that has only  one asymptotically flat end (fig. \ref{Wormhole}(b)).  The ability to do this is
 based on the fact that far from a wormhole mouth,  spacetime is  approximately flat. Usually spacetime is only
approximately flat far from a wormhole, in which case
 an exact solution in a world with two asymptotically flat ends leads only to an approximate solution in a world with a single asymptotically flat end.
 However, in the present context,
we can assume that the function $r(u)$ in fig. \ref{FourChoices}(c) is identically equal to $u$ for $|u|>c$ (for some constant $c$); then 
the metric in fig. \ref{Wormhole}(a) is identically equal to the standard Euclidean metric on $\R^D$ outside of a compact set on each branch.   Given this, 
the cut and paste procedure required to get a wormhole metric of the sort sketched in fig. \ref{Wormhole}(b) is exact.
So we get complex solutions of Einstein's equations of wormhole type, with vanishing action, on a spacetime that at infinity reduces to $\R^D$ with its standard flat metric.

Finally, we can consider the case that the curve $r(u)$ is a circle embedded, or at least immersed, in the $r$-plane minus
the point $r=0$.   In this case, we get a flat complex metric on $\S^1\times \S^{D-1}$, again with zero action.

Somewhat similarly, we can make exotic complex solutions of Einstein's equations with a positive cosmological constant
starting with the standard metric on a round sphere of radius $\rho$:
\be\label{startround} \d s^2=\rho^2( \d \theta^2+\cos^2\theta \d\Omega^2). \ee
One important difference from the flat case is that the action $I$ of a compact solution of Einstein's equation with cosmological constant does not
vanish; it is a negative multiple of the volume $V$, 
\be\label{multiple} I=-\frac{2}{D-2} \Lambda V,  \ee where $\Lambda$ is the cosmological constant.
After solving Einstein's equations to determine $\rho$ and therefore $V$ in terms of Newton's constant $G$ and $\Lambda$, one finds that 
 $\Lambda V\sim 1/G^{D/2}\Lambda^{(D-2)/2}$.    Thus the action  is large and negative if  $\Lambda$ is small and positive, a fact that has
 been offered as the reason that
 ``the cosmological constant is probably zero'' \cite{HawkingProbably}.    The de Sitter entropy is defined as $S=-I$.   

Rather as before, we can construct complex metrics that solve the Einstein equations  with a cosmological constant by considering curves in the complex $\theta$ plane.
The resulting metric has the form $\d s^2= \rho^2(\theta'(u)^2\d u^2 +\cos^2\theta(u)\d\Omega^2)$.
First let us discuss compact solutions.   The function $\cos^2 \theta$ vanishes at $\theta=(n+1/2)\pi, $ $n\in\Z$.   A curve connecting two of these zeroes, and otherwise avoiding
all zeroes, will
give a manifold that is topologically $\S^D$.   Of course, there are many homotopy classes of such embedded or immersed curves, even after fixing the endpoints.
 The most basic example is a  straight line on the real $\theta$ axis between two consecutive zeroes.  This gives the standard real metric
on a round sphere.   Curves between nonconsecutive zeroes give exotic complex solutions of the Einstein equations. 
The volume is defined in Riemannian geometry as $\int \d^Dx \sqrt {\det g}$.  In the case of a complex invertible metric, it is not immediately obvious
what sign one should take for $\sqrt{\det g}$, and on a manifold that is not simply-connected, in general there
can be an inconsistency in defining this sign.   For the allowable metrics of Kontsevich and Segal, which we discuss in section \ref{allowable}, there is a natural
choice of sign.   In the present section, we will not try to be precise about the sign of the volume.   

For a metric derived from the round metric (\ref{startround}) by a choice of curve $\theta(u)$, the volume can be computed as the integral of the differential
form $\Psi= \rho^D \cos^{D-1}\theta \,\d \theta\d\Omega$. (In this formulation, the sign ambiguity appears when one picks an orientation of the cycle on which one wishes
to integrate this differential form.)   
   Hence, letting $v_D$
denote the volume of a standard $D$-sphere of unit radius, the volume for a solution based on a curve from $(m+1/2)\pi$ to $(n+1/2)\pi$ is
\be\label{volform} V_{m,n}=\rho^D v_{D-1} \int_{(m+1/2)\pi}^{(n+1/2)\pi}\d \theta \,\cos^{D-1} \theta. \ee
By Cauchy's theorem, this integral depends only on the endpoints of the curve, not on the path taken between those endpoints.  That happened because
the differential form $\Psi$ is closed, and means that the volume depends only on the homology class of the curve $\theta(u)$.  
For even $D$, $V_{m,n}$  vanishes if $n-m$ is even, and equals $v_D\sign(n-m) $ if $n-m$ is odd.   For odd $D$,  $V_{m,n}=(n-m) v_D$.    Thus,
for odd $D$, there exist complex solutions of Einstein's equations with topology $\S^D$ and action much more negative than the action of the standard
real solution.   

Another basic example is the  curve $\theta=\i u$, with $-\infty<u<\infty$.  The resulting line element
\be\label{lorentz}\d s^2=\rho^2(-\d u^2+\cosh^2u \,\d\Omega^2) \ee
describes de Sitter spacetime of Lorentz signature with radius of curvature $\rho$.    An embedded or immersed curve from $\theta=-\i\infty$ to $\theta=+\i\infty$ that avoids
 the zeroes of $\cos \theta$ gives a complex metric that is asymptotic to de Sitter space in the far past and future.   
 
 Now suppose that we want a solution of the complex Einstein equations that has no boundary to the past, and coincides with de Sitter space
 in the future.   Such solutions have been discussed in the context of the ``no boundary'' wavefunction of the universe \cite{HH}.   We can get such a solution
 from a curve $\theta(u)$ with $u$ running over the half-line $[0,\infty)$, provided $\theta(0)$ is one of the zeroes of $\cos\theta$, and $\theta(u)=\i u$ for large $u$.
 The case most often discussed is a curve that goes along the real axis from $\theta=-\pi/2$ (or $\theta=\pi/2$) to $\theta=0$ and then continues along the imaginary axis
 from $\theta=0$ to $\theta=\i\infty$ (or to some given point on the imaginary $\theta$ axis).    This is usually described as follows.   The path integral along the real axis from $\theta=\pm \pi/2$ to $\theta=0$  is a Euclidean path
 integral that prepares an initial state; then the path integral along the imaginary $\theta$ axis  up to a point $\theta=\i u$ is a Lorentz signature path integral that propagates the state for an
 arbitrary real time $\rho u$.   The Euclidean part of the contour contributes an action $-S/2$ (where again $S$ is the de Sitter entropy) and the Lorentz
 signature part of the contour makes an imaginary contribution 
 \be\label{imcont}-\i \I(u) = - \i   \frac{\rho^D c_D v_{D-1}}{G^{D/2} \Lambda^{(D-2)/2}} \int_0^u \d u'  \,\cosh^{D-1} u' .\ee
In a classical approximation,  the time-dependent state created by the path integral on this contour is described by the exponential of the action:
$e^{-I}=e^{S/2} e^{\i \I(u)}$.    The
 real factor $e^{S/2}$ is the norm of the state.     The oscillatory factor $e^{\i \I(u)}$ describes the real time evolution of de Sitter space,
 as discussed for example in \cite{HalHar}.    
 
 However, in the world of complex metrics, we can easily construct additional complex metrics that could conceivably represent the creation of de Sitter
 space from nothing, in the context of the no boundary wavefunction.   With $u$ still ranging over the half-line $[0,\infty)$, we can choose $\theta(0)$ to
 be a zero of $\cos r$ at $r=-(n+1/2)\pi$, for any $n$, while keeping $\theta(u)=\i u$ for large $u$.  
  For odd $D$, this multiplies the real part of the action by $2n+1$ and gives
 a wavefunction proportional to $e^{(n+1/2)S}e^{\i \I(u)}$.    Thus, naively, we can increase the amplitude to ``create a universe from nothing'' by increasing $n$.
 
By taking the curve $\theta(u)$ to be a circle, one can similarly get complex metrics on $\S^1\times \S^{D-1}$ with 
zero volume that satisfy Einstein's equations with a cosmological constant.

What hopefully stands out from this discussion is that many or all of the exotic examples are going to give unphysical results if included in path integrals.  
  One way or another, they must be excluded.   
  With this in mind, after describing in section \ref{allowable} the allowable metrics of Kontsevich and Segal, we will show that the exotic
 examples considered in this section are not allowable.

\section{Allowable Metrics}\label{allowable}

On a spacetime $M$ of dimension $D$ with a complex invertible metric $g$, consider a $p$-form gauge field $A$ with $p+1$-form field strength $F=\d A$; set
$q=p+1$.  The usual action is
\be\label{usact}I_{q} =\frac{1}{2 q!}\int_M \d^D x \sqrt{\det g} g^{i_1 j_1}\cdots  g^{i_{q} j_{q} }F_{i_1 i_2 \cdots i_{q}} F_{j_1 j_2\cdots j_{q}}. \ee
The metric $g$ is {\it allowable}, in the sense of Kontsevich and Segal, if $I_q$ has positive real part for every nonzero (real) $q$-form $F$, for any $0\leq q\leq D$.
This amounts to a pointwise condition:\footnote{For $q\leq 2$, this condition was proposed by Louko and Sorkin
in footnote 8 of \cite{SL} (see also their discussion of eqn. (2.20)).   Note that the case $q\leq 2$ suffices in $D=2$, which is the main case considered
in \cite{SL}.    One might think that one could weaken the condition (\ref{nsact}) by
taking advantage of the fact that $F=\d A$ obeys a Bianchi identity.   This is actually not true, because every $q$-form $F$ can be written as a sum $F_1+\star F_2$,
where $F_1$ is a closed $q$-form and $F_2$ is a closed $(D-q)$-form ($\star$ is the Hodge star). Condition (\ref{usact}) for closed $F_1$ and 
$F_2$ implies (\ref{nsact})  for arbitrary $F$.}
\be\label{nsact} \Re\left( \sqrt{\det g} g^{i_1 j_1}\cdots  g^{i_{q} j_{q} }F_{i_1 i_2 \cdots i_{q}} F_{j_1 j_2\cdots j_{q}}\right) >0,~~0\leq q\leq D, \ee
for any real, nonzero $q$-form $F$.

A motivation for imposing this positivity is that it makes the path integral of a $p$-form gauge field convergent, for any $p$.    The idea in \cite{KS}  is
that quantum field theory in general -- not just the free theory of a $p$-form field -- is well-defined for a general allowable metric.   More specifically, the hope
is that this property can substitute for at least some of the standard axioms of quantum field theory.    Some evidence in this direction is given in \cite{KS}.
The  condition (\ref{nsact})  is imposed for all $0\leq q\leq D$, although the motivation in terms of the field strength of a $q-1$-form field does not apply for $q=0$.   
The $q=0$ condition is just $\Re\,\sqrt{\det g}>0$.   One way to motivate the $q=0$ case of the condition (\ref{nsact}) is to observe that a zero-form field $\phi$
is a scalar field, which could have a bare mass $m$.   Positivity for real $\phi$ of the real part of the corresponding action $\frac{m^2}{2}\int_M\d^Dx\sqrt{\det g} \phi^2$ gives the $q=0$ case of eqn. (\ref{nsact}). 

Many questions can be asked about whether the condition of allowability is either necessary or sufficient for well-definedness of quantum field theory.   We will
be rather brief with such questions, as it will not be possible to resolve them definitively.   In terms of sufficiency, one can ask whether well-definedness
of $p$-form theories for all $p$  in a spacetime with a given complex metric 
is sufficient to ensure well-definedness of general quantum field theories in that spcaetime.   Here it is worth noting that free field theories of
massless bosonic fields\footnote{Bare masses do not help; it is difficult to construct consistent  couplings to gravity of massive fields other than $p$-form fields 
except via Kaluza-Klein theory or string theory \cite{Nappi,NW}.} 
other than $p$-forms do exist, but those theories do not have gauge-invariant stress tensors \cite{WW} and therefore cannot be defined in curved
spacetime.   Moreover,  non-free ultraviolet-complete theories of massless fields other than $p$-forms 
are not known, even in flat spacetime.    So $p$-form theories are actually important examples of quantum field
theories.   For $p=0,1$, there are nonlinear versions of $p$-form theories -- nonabelian gauge theory for $p=1$ and nonlinear sigma-models for $p=0$ -- and
for all $p$, there are mildly nonlinear theories in which, for example,  the field strength of a $p$-form field $A$ is not $\d A $ but $\d A+B\wedge C$, where $B,C$
are forms of degree $r,s$ with $r+s=p+1$.   In any of these cases, the real part of the action is positive in an allowable metric.  So it is plausible that 
known quantum field theories that are associated to underlying classical theories can all be consistently coupled to an allowable metric.    

Concerning necessity,
one might ask if the condition of allowability is unnecessarily strong.   For example, let $g$ be an allowable metric and let $g_\varphi
=e^{\i\varphi}g$ for some real $\varphi$.   Replacing $g$ by $g_\varphi$ would multiply the action (\ref{usact}) by a factor $e^{\i\varphi(D/2-q)}$, potentially
spoiling the positivity of $\Re \,I_q$.  Can we compensate by rotating the integration contour for the $p$-form path integral by a phase, $A\to A e^{-\i\varphi(D/4-q/2)}$?
This might make sense for perturbative fluctuations, but as noted in \cite{KS}, the partition function of a $p$-form field also involves a sum over quantized integer
fluxes that cannot be rotated in the complex plane. So some condition along the lines of allowability is needed, though from this point of view a weaker condition
might suffice.   Another important point is that one may want the path integral for a quantum field theory in curved spacetime to have a Hilbert space interpretation.
For this, the path integral of the matter fields has to be defined by local considerations, not by a completely general analytic continuation.  
Yet another issue is that the Wick rotation of the matter fields
in general may multiply the path integral measure by an ill-defined phase.

In this article, we will not try to address such issues and instead will 
concentrate on the following two questions.  Does a restriction to allowable metrics remove unwanted examples such as the solutions of Einstein's equations
discussed in section \ref{examples}?   And where useful results have come from a consideration of complex solutions of Einstein's equations, have the
metrics in question been allowable?   We address the first question here, after summarizing some additional facts from \cite{KS}, and we explore the second
question in section \ref{useful}.

A simple and useful characterization of allowable metrics was found in \cite{KS}.  The $q=1$ case of eqn. (\ref{nsact}) tells us that the real part of
the matrix $W=\sqrt{\det  g}\, g^{ij}$ is positive definite.   Writing $W=A+\i B$, where $A$ and $B$ are real, it follows that $A$ and $B$ can be simultaneously
diagonalized by a suitable choice of real basis.   (First one picks a basis to put $A$ in the form $\delta_{ij}$; such a basis is unique up to an orthogonal transformation,
which can be used to also diagonalize $B$.)   In such a basis, $W$ is diagonal, and therefore so is $W^{-1}=g/\sqrt{\det g}$.   Multiplying by the scalar
$\sqrt {\det g}$, it follows that $g$ is diagonal in this basis:
\be\label{diagbasis} g_{ij}= \lambda_i \delta_{ij},~~i,j=1,\cdots, D. \ee
Therefore $\sqrt {\det g}=\prod_i \sqrt\lambda_i$.   Condition (\ref{nsact}) for $q=0$ tells us to pick the sign of the square root so that $\Re\,\sqrt{\det g}>0$.
Condition (\ref{nsact}) now says that for any subset\footnote{\label{duality} If $x$ is a complex number, then $\Re\,x>0$ if and only if $\Re\,1/x>0$.   Using this,
one can see that condition (\ref{prods}) for a given set $S$ is equivalent to the same condition for the complement of $S$.  Hence it suffices to consider
sets $S$ of cardinality at most $D/2$.    Equivalently, it suffices in this
construction to consider $p$-form fields with $p\leq D/2-1$.   This is related to the duality in $D$ dimensions between a $p$-form field and
a $(D-2-p)$-form field.}
$S$ of the set $\{1,2,\cdots,D\}$, 
\be\label{prods} \Re\left({\sqrt {\det g}}\prod_{i\in S} \lambda_i ^{-1}\right)>0.\ee
This holds precisely if 
\be\label{mineq}\sum_{i=1}^D |\mathrm{Arg}\,\lambda_i|<\pi.\ee This statement is Theorem 2.2 in \cite{KS}.

For example, a Lorentz signature metric $\d s^2=-\d x_1^2+\sum_{j=2}^D \d x_j^2$ is not allowable, since this corresponds to the case $\Arg\,\lambda_1=\pi$.
The inequality (\ref{mineq}) is just barely violated, so a Lorentz signature metric is on the boundary of the space of allowable metrics.  In fact, a Lorentz
signature metric is on the boundary of the space of allowable metrics in two different ways, since for positive $\epsilon$, either of the two metrics
\be\label{eitheror}\d s_\pm^2=-(1\mp \i\epsilon)\d x_1^2+\sum_{j=2}^D \d x_j^2 \ee
is allowable.   The difference between the two cases involves the sign of $\sqrt {\det g}$.   Since we are instructed in eqn. (\ref{nsact}) to choose the sign
of the square root such that $\Re\,\sqrt{\det g}>0$, it follows that for $\epsilon\to 0$, $\sqrt{\det g}$ approaches the positive or negative imaginary axis depending on
the sign in eqn. (\ref{eitheror}). 
Therefore, the sign of the Lorentz signature action $\int\d^D x \sqrt g \L$ (where $\L$ is the Lagrangian density) depends on the sign of the $\pm\i\epsilon$ term.
The choice $\epsilon>0$ leads to the standard Feynman integral computing real time propagation by $\exp(-\i H t -\epsilon H)$, where $H$ is the Hamiltonian and
$\epsilon$ appears in  the usual
Feynman $\i\epsilon$, and the choice $\epsilon<0$  leads to a complex conjugate Feynman integral that computes $\exp(+\i H t-\epsilon H)$.   
In the Schwinger-Keldysh approach
to thermal physics, one sign leads to propagation of the ket vector and the other sign leads to propagation of the bra.    Which is which is a matter of convention.
Clearly, the two metrics in eqn. (\ref{eitheror}) cannot be considered ``close,''  even for small $\epsilon$, and one is definitely not allowed to interpolate
between them by letting $\epsilon$ change sign.

Perhaps it is worth stressing that the ability to regularize in this way a Lorentz signature metric as an allowable complex metric does not depend at all on the causal
properties of the Lorentz signature metric.    So from this point of view, it is perfectly sensible to consider Lorentz signature spacetimes with closed timelike curves.

The criterion (\ref{mineq}) implies, as was explained in \cite{KS}, that the space of allowable complex metrics is contractible
onto the space of Euclidean metrics.\footnote{The space of Euclidean metrics in turn is contractible to a point, by a standard argument. 
 If $g_0$ is some chosen Euclidean signature metric on $M$
and $g$ is any other such metric, then for $0\leq t\leq 1$, $g_t=(1-t)g + t g_0$ is a Euclidean signature metric.    Here we use the fact that the only constraint
on a real symmetric tensor to make it a Euclidean signature metric is that it should be positive-definite; if $g$ and $g_0$ have this property, then so does $g_t$.
By letting $t$ vary from 0 to 1, we contract the space of all Euclidean signature metrics on $M$ onto the metric $g_0$, showing that the space of Euclidean
signature metrics is contractible.  Note that this argument is not valid (and the conclusion is not true) for real metrics of Lorentz signature, or any signature other
than Euclidean signature. }  Indeed, eqn. (\ref{mineq}) implies that for all $i$, 
$\lambda_i$ is not on the negative real axis, so there is a canonical path to rotate $\lambda_i$ to the positive
real axis while always satisfying the condition (\ref{mineq}): one rotates $\lambda_i$ in the upper half plane if $\Im\,\lambda_i>0$, and in the lower half plane
if $\Im\,\lambda_i<0$.    It follows that any topological invariant that can be defined using an invertible metric, such as the integrals that define the Euler characteristic
or the Pontryagin numbers of $M$, takes the same value for an allowable complex metric as for a Euclidean metric.    It was shown in \cite{SL}, with the example
of the Gauss-Bonnet integral in two dimensions, that in general this is not true for complex invertible metrics.   We return to this point in Appendix \ref{gb}.

The $q=0$ case of eqn. (\ref{nsact}) implies that if $M$ is a manifold with allowable metric, then its volume $\int_M\d^D x \sqrt g$ has positive real part.
In \cite{KS}, it is shown that eqn. (\ref{mineq}) implies that if $M$ has an allowable complex metric, then the induced metric on any submanifold $N$ of $M$
is also allowable.  Hence the volume of any such $N$ has positive real part.  One can take this as an indication that perturbative strings and branes
make sense on a manifold with allowable complex metric.    (In the case of perturbative string theory, one will need to impose worldsheet conformal invariance,
as in the more familiar case of a Euclidean metric on $M$.)   

Now we will use eqn. (\ref{mineq}) to show that the problematic examples of section \ref{examples} are not allowable,.   First consider
the flat metrics of the form $\d s^2= r'(u)^2\d u^2+r(u)^2\d\Omega^2$, where $r(u)$ is a curve in the complex plane.    If there is any value of $u$
for which $r(u)$ is imaginary, then at that value of $u$, $r(u)^2 \d\Omega^2$ is negative-definite, so $D-1$ of the $\lambda_i$ in eqn. (\ref{diagbasis}) are negative
and eqn. (\ref{mineq}) is not satisfied at that value of $u$.    If $r(u)$ is never imaginary, then (except for possible endpoints at $r(u)=0$) the curve is contained in one of the half-planes $\Re\, r(u)>0$
and $\Re\, r(u)<0$.   If the curve is homotopic to the positive or negative $u$ axis, the corresponding metric is equivalent to the standard flat metric on $\R^D$.
Otherwise,
  there is some value of $u$ at which $\Re\,r(u)$ has a maximum
or a minimum.   At such a point, $\Re\,r'(u)=0$ so $r'(u)$ is imaginary.
Hence $r'(u)^2\d u^2$ is negative definite and one of the $\lambda_i$ in eqn. (\ref{diagbasis}) is negative, contradicting eqn. (\ref{mineq}).   The behavior
near a maximum or minimum of $\Re\, r(u)$ actually consists of a forbidden transition between the two choices of sign in eqn. (\ref{eitheror}).

Finally, consider the metrics $\d s^2= \rho^2(\theta'(u)^2\d u^2 +\cos^2\theta(u)\d\Omega^2)$ that satisfy Einstein's equations with a cosmological constant.
Here $\theta(u)$ is a curve that avoids the points $\theta=\pi(n+1/2)$, $n\in\Z$, except at  endpoints.
For an allowable metric, $\Re\,\cos\theta(u)$ must be nonzero, except possibly at endpoints; otherwise $\cos^2\theta(u)\d\Omega^2$ is negative
definite and eqn. (\ref{mineq}) is violated.     To avoid vanishing of $\Re\,\cos\theta(u)$, 
the curve $\theta(u)$ must be confined to a strip $(n-1/2)\pi\leq \Re\,\theta\leq ( n+1/2)\pi$, for some $n\in \Z$.   This excludes many of the exotic possibilities
described in section \ref{examples}, including the closed universes and the solutions describing ``creation of a universe from nothing'' that have action more negative than
the standard values.   The other exotic possibilities are excluded by the fact that $\Re\,\theta(u)$ cannot have a maximum or minimum along the curve,
since at such a maximum or minimum, $\theta'(u)^2\d u^2$ is negative definite.

\section{Some Useful Complex Saddles}\label{useful}

 \begin{figure}
 \begin{center}
   \includegraphics[width=2.9in]{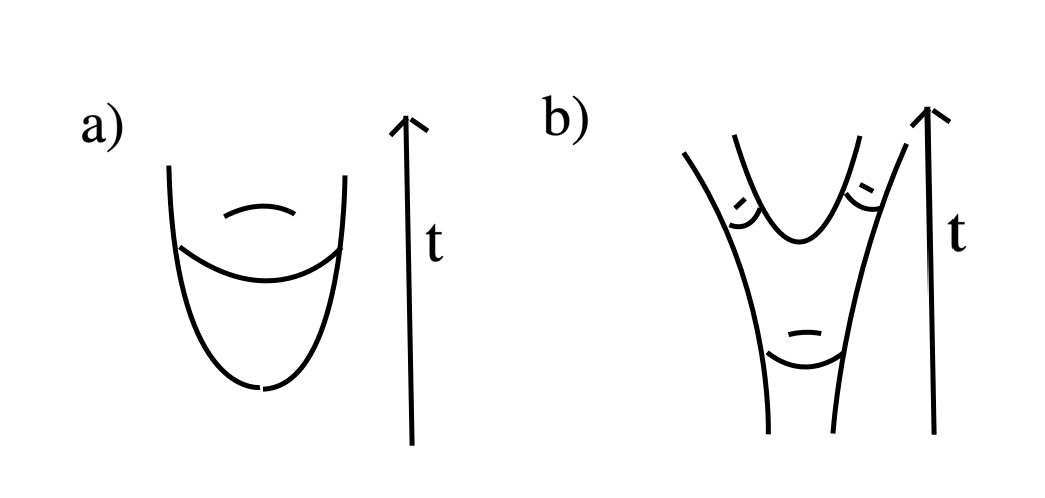}
 \end{center} 
\caption{\small   (a) Creation of a universe from nothing. (b) Splitting of a universe in two.    In each case the time coordinate $t$ runs vertically, as shown. \label{Example}}
\end{figure} 

Our goal in this section is to examine some examples in which results that appear to be physically sensible have been obtained by considering
complex metrics on spacetime.   We will see in these examples that the metrics in question are allowable.  However, we consider only a few examples and it is not clear what conclusions can be drawn.\footnote{There are also proposals in the literature
for applications of some non-allowable metrics.   See for example \cite{MTZ}.}

\subsection{Topology Change In Lorentz Signature}\label{topchange}

The first example that we will consider is topology change in Lorentz signature, which was considered originally by Louko and Sorkin \cite{SL} from a similar
point of view.    See for
example \cite{DS,Dowker,Borde,Sor2} for further discussion of topology change in real time.

  For closed universes in two spacetime dimensions, the basic examples of topology change are the creation of a closed universe from
nothing (fig. \ref{Example}(a)), the splitting of a closed universe in two (fig. \ref{Example}(b)), and the time reverses of these.   Neither is possible with
a smooth and everywhere nondegenerate Lorentz signature metric.     However, it is possible to pick a Lorentz signature metric which is smooth,
and is nondegenerate except at one point in spacetime, at which the topology change occurs.   For example, on a spacetime
that describes creation of a closed universe from nothing, we can take real coordinates $x,y$ such that the creation event occurs at $x=y=0$, and consider
the line element
\be\label{bogo} (x^2+y^2)(\d x^2+\d y^2)-\zeta (x\d x + y\d y)^2\ee with a constant $\zeta$.  The metric (\ref{bogo}) smooth, and is nondegenerate except at $x=y=0$.
     At $x=y=0$, the metric is degenerate -- it vanishes.  Provided that $\zeta>1$, this metric has  Lorentz signature except at $x=y=0$; $t=x^2+y^2$ can be viewed as a ``time'' coordinate, and the metric describes a circle
that is ``created'' at $t=0$ and grows in proportion to $ t$.
 
   Some regularization
is required, since presumably the coupling of quantum fields to a degenerate metric is not well-defined.
One way to regularize the metric is to replace the line element with\footnote{Louko and Sorkin used a slightly different regulator that is also consistent
with eqn. (\ref{mineq}) in a suitable range of $\zeta$.} 
\be\label{ogo} (x^2+y^2+\tilde\epsilon(x,y))(\d x^2+\d y^2) -(\zeta\pm\i\epsilon)(x\d x+y\d y)^2. \ee
In the framework discussed in the present paper, $\epsilon$ should be nonzero for all $x,y$ because a Lorentz signature metric is regarded as a limiting
case of a complex invertible metric that satisfies eqn. (\ref{mineq}).   The role of $\tilde\epsilon$ is to make the metric nondegenerate for all $x,y$; for this,
$\tilde\epsilon$ should be positive at $x=y=0$ but can vanish except very near that point.   To see that the line element (\ref{ogo}) corresponds to
an allowable metric, let $V$ be the one-form $x\d x+y\d y$, and let $W$ be a one-form that is orthogonal to $V$.   Then for $(x,y)\not=(0,0)$,  (\ref{ogo}) has the general
form
\be\label{pogo} -(A\pm\i\epsilon)V\otimes V +B W\otimes W,~~A,B>0.\ee
This is manifestly consistent with eqn. (\ref{mineq}).   At $x=y=0$, the metric has Euclidean signature and again eqn. (\ref{mineq}) is satisfied.

Splitting of a closed universe into two can be treated similarly.  One can pick a ``time'' coordinate $t$ whose differential is nonzero except at an isolated
saddle point, at which the topology change occurs (fig. \ref{Example}(b)).  Near the saddle point, one can pick local coordinates $x,y$ with $t=x^2-y^2$.
The line element
$(x^2+y^2)(\d x^2+\d y^2)-\zeta (x \d x - y\d y)^2$, $\zeta>1$
describes
a smooth metric that has Lorentz signature everywhere except at the saddle point, where it vanishes.    This again can be regularized to give an allowable complex metric 
$(x^2+y^2+\tilde\epsilon(x,y))(\d x^2+\d y^2)-(\zeta\pm\i \epsilon )
(x \d x - y\d y)^2$.

\begin{figure}
 \begin{center}
   \includegraphics[width=2in]{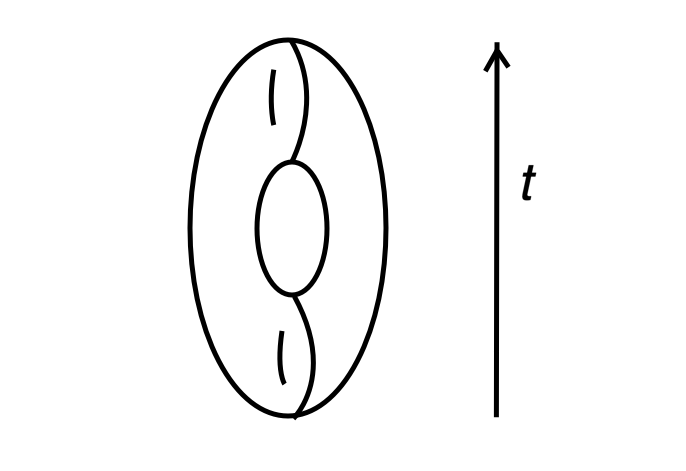}
 \end{center} 
\caption{\small  A torus embedded in $\R^3$ in such a way that the coordinate $t$, running vertically, has a maximum, a minimum, and two saddle points.
An analogous embedding of a surface of genus $g$ has a maximum, a minimum, and $2g$ saddle points.  \label{Torus}}
\end{figure}

Now consider the case that spacetime is a closed two-manifold $\Sigma$ of genus $g$.  The Gauss-Bonnet integral 
$I_\Sigma=\frac{1}{4\pi}\int_\Sigma \d^2x \sqrt g{R}$ appears 
in the action of ``Euclidean'' quantum gravity with a negative coefficient.  If we pick on $\Sigma$ a Euclidean signature metric   $h_{ab}$, then the Gauss-Bonnet
theorem gives $I_\Sigma=2-2g$.   On the other hand, it is possible to pick on $\Sigma$ a Morse function $t$ which has one local minimum, one local maximum,
and $2g$ saddle points (fig. \ref{Torus}).  Together these points are called the critical points of $t$.  As noted by Louko and Sorkin \cite{SL}, given $h_{ab}$ and $t$,
one can construct the metric
\be\label{themetro} g_{ab}=h_{ab}( h^{cd}\partial_c t \partial_d t )-\zeta \partial_a t \partial_b t,\ee
which is everywhere smooth, and has Lorentz signature except at the critical points, where it vanishes.   A simple regularization that gives an allowable
complex metric is to replace $g_{ab}$ with
\be\label{themetric}\tilde  g_{ab}=h_{ab}( h^{cd}\partial_c t \partial_d t +\tilde \epsilon)-(\zeta\pm\i\epsilon) \partial_a t \partial_b t,\ee
with  $\tilde\epsilon>0$ near critical points (but potentially vanishing except near critical points) and $\epsilon>0$.
Now let us consider the Gauss-Bonnet integral $I_\Sigma$ for this kind of metric.    As noted by Louko and Sorkin, $ \sqrt g R$ is imaginary for a Lorentz signature
metric, and therefore in the limit $\epsilon,\tilde \epsilon\to 0$, the expected real contribution $2-2g$ from $I_\Sigma$ must be localized at the critical points of the
function $t$.   Indeed, they showed that a local maximum or minimum of $t$ contributes $1$ to $I_\Sigma$, while a critical point contributes $-1$.    As was explained
in section \ref{allowable}, the Gauss-Bonnet integral has its standard value for an allowable complex metric.   This is not so for a general complex metric,
as observed by Louko and Sorkin and further discussed in Appendix \ref{gb}.

Topology change in Lorentz signature in any dimension $D$ can be treated similarly.   On any manifold $M$, one can choose a ``Morse function'' $t$ that has only
isolated, nondegenerate critical points.    A metric of the form (\ref{themetric}) is smooth and has Lorentz signature except at the critical points, where it vanishes.
The regularization (\ref{themetric}) makes sense in any dimension and gives a complex allowable metric.

\subsection{The Hartle-Hawking Wavefunction}\label{hhw}

An important application of complex spacetime metrics is to the Hartle-Hawking wavefunction of the universe \cite{HH}.   In $D$ dimensions, one
considers a $D-1$-dimensional manifold $Y$ with metric $g_{D-1}$.   The Hartle-Hawking wavefunction $\Psi_{HH}(g_{D-1})$ is formally defined
by a sum over all manifolds $M$ with boundary $Y$, with the contribution of each $M$ to the sum being the gravitational path integral over
metrics on $M$ that restrict on the boundary to $g_{D-1}$.   

A saddle point in this context is a classical solution of the Einstein equations on $M$ that restricts to $g_{D-1}$ on $Y$.   This problem makes sense
for Euclidean metrics.
If the Einstein-Hilbert action in Euclidean signature were bounded below, one might hope that real saddle points would exist -- a metric that minimizes
the action for given $g_{D-1}$ would be an example.  Such a real saddle point might fail to exist if when we try to minimize
the action, $M$ develops
a singularity.   The analog of this actually happens for instantons in Yang-Mills theory with Higgs fields on $\R^4$; in an instanton sector, there is no true classical solution,
since the action can be reduced by letting the instanton shrink to a point.  Similar behavior can occur in gravity; see for example \cite{SSSJT}. 
Even if a classical minimizer does not exist, if the action were bounded below, it would have a greatest lower bound $\J(g_{D-1})$ and the asymptotic
behavior of the Hartle-Hawking wavefunction, near a classical limit or when $g_{D-1}$ describes a manifold of large volume,
 would be $\Psi_{HH}(g_{D-1})\sim \exp(-\J(g_{D-1}))$. If the greatest lower bound on the Euclidean
  classical action is not achieved by any smooth classical field (because a singularity develops when one tries to minimize the action), but can be approximated
 by a sequence of smooth classical fields, one could
 describe this roughly by saying that there is no true saddle point but there is a 
 virtual saddle point at infinity in field space, analogous to a point instanton in Yang-Mills theory.

  \begin{figure}
 \begin{center}
   \includegraphics[width=2.9in]{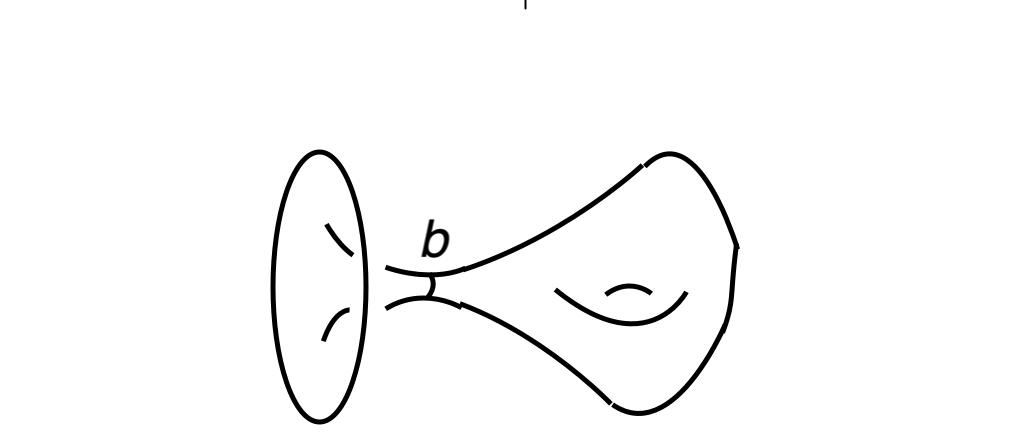}
 \end{center}
\caption{\small  The genus 1 contribution to the one-boundary partition function in JT gravity with a negative cosmological constant, as studied 
in \cite{SSSJT}.    There is no conventional critical point; in searching for one, one finds that the length $b$ of the indicated geodesic that separates
the disc from the genus 1 surface tends to shrink to 0, because of a contribution to the action proportional to $b^2$.   Instead of a conventional critical
point there is a critical point at infinity -- or more precisely, at $b=0$.   This critical point describes a hyperbolic disc with one cusp and a hyperbolic torus with one cusp,
with the cusp points identified. It controls the semiclassical behavior of the path integral. \label{gluing}}
\end{figure} 
 
In many known examples, the gravitational path integral in the semiclassical limit is dominated by a critical point at infinity
rather than a conventional critical point; for one such example, see fig. \ref{gluing}.   In all cases in which any candidate has been suggested for the semiclassical
behavior of a gravitational path integral, it is dominated by a conventional critical point or a critical point at infinity. 
The reason for this is just that if a critical point, possibly at infinity, has not been found, then the asymptotic behavior of the gravitational path integral
is unknown.
 From the standpoint of a possible ultraviolet
completion of Einstein gravity, there is potentially no fundamental difference between an ordinary critical point and a critical point at infinity.
What looks in one description like a singular configuration might be perfectly smooth in a more complete description.
Precisely this has happened to many types of singularity in gauge theory and gravity that have turned out to be perfectly smooth from the vantage point of string theory.
In Morse theory, where topological meaning is given to the critical points of a function, it is a standard fact
that on a non-compact manifold in general one has to include critical points at infinity.  So their appearance in the gravitational
path integral, where one deals with critical points of the action functional
on the noncompact space of all real or complex metrics on a manifold $M$, should not be too much of a surprise.

  In fact, however, the Euclidean action for gravity is unbounded below, and in general the boundary value problem associated to the Hartle-Hawking wavefunction
  is believed to have no classical solution of Euclidean signature (not even a virtual solution at infinity).  For instance, 
 consider Einstein's equations with positive cosmological constant $\Lambda$ and choose $Y$ to be a sphere
with a round metric with a very large radius of curvature (compared to the length scale set by $\Lambda$). It is believed  there  is no real classical solution, not even in a
limiting sense.

As is explained in \cite{HalHar}, this phenomenon is actually a necessary condition for the Hartle-Hawking wavefunction of the universe to make sense in the
context of cosmology.
If the Einstein-Hilbert action in Euclidean signature were positive-definite like the action of a conventional scalar field, then the wavefunction $\Psi_{HH}(g_{D-1})$ would
behave semiclassically as $\Psi_{HH}(g_{D-1})\sim \exp(-\J(g_{D-1}))$, as noted earlier, where $\J$ is the greatest lower bound on the action.   If gravity
had similar positivity properties to scalar field theory, where the action is strictly positive unless the field is constant, we would expect $\J$ to be strictly positive
except in very special cases.
By a simple scaling argument, $\J$ would grow if the metric on $Y$ is scaled up to large volume, and hence the Hartle-Hawking
wavefunction would vanish exponentially when $Y$ is large.   This exponential decay of the wavefunction for large volumes would be analogous
 to the fact that, for a real scalar field $\phi$, a wavefunction $\Psi(\phi_{D-1})$ defined similarly to the Hartle-Hawking wavefunction vanishes exponentially  for large $\phi$.
 The scalar  analog  of the Hartle-Hawking wavefunction is defined by choosing a particular $M$ with boundary $Y$, and performing a
 path integral over $\phi$ fields on $M$ that restrict to $\phi_{D-1}$ on $Y$.   This wavefunction vanishes exponentially for large $\phi$ because the usual action of a scalar
 field is positive-definite, and grows  when the boundary values are increased. In fact, the action grows quadratically with $\phi$ so the wavefunction vanishes
 as the exponential of $\phi^2$.

In the case of gravity, the Hartle-Hawking wavefunction can behave differently because, as the Euclidean action is not bounded below,
it is possible for there to be no real critical point (even at infinity), and the integral can potentially be dominated by  complex critical points.
For instance, in the example of the round sphere of large radius in a world with 
$\Lambda>0$, though there is no real saddle point, there are complex saddle points that give an oscillatory contribution to the path integral. 
Moreover, this contribution can be large near a classical limit, because the real part of the ``Euclidean'' action can be negative.
  As explained in the 
discussion of eqn. (\ref{startround}), there are both conventional complex saddles, which involve trajectories from $\theta=\pm \pi/2$ to a point $\theta=\i u$ on the
imaginary axis, and unconventional ones, which start at $\theta=\pm (n+1/2)\pi$, $n>0$.    The unconventional ones lead to apparently unphysical behavior, but
we noted at the end of section \ref{allowable} that they are not allowable.   An equally important fact is that the conventional saddle points, starting at $\theta=\pm \pi/2$,
are allowable.  To get an allowable metric, we can start with the usual idea of a straight line from $\theta=\pi/2$ to $\theta=0$ joined to a straight line from $\theta=0$
to $\theta=\i u$, and modify this slightly to get a smooth path along which $\Re\,\theta'(u)$ is everywhere nonzero.   This gives the allowable metric
 $\d s^2=\rho^2(\theta'(u)^2\d u^2 +\cos^2 \theta(u)\d \Omega^2)$.

As many authors have pointed out, although the exponentially large amplitude for this allowable trajectory that ``creates a universe from nothing'' is interesting,
it cannot be the whole story for cosmology, since this mechanism would tend to produce an empty universe with the smallest possible positive value of the cosmological
constant.   For an alternative treatment of these solutions, based on a different boundary condition in which the second fundamental form of the boundary is fixed,
rather than the metric of the boundary, see \cite{BH}.  

\subsection{Timefolds}\label{timefolds}

In many contexts, for instance in real time thermal physics and in various analyses of gravitational entropy, such as \cite{HRT,MM,CDMRW}, it is convenient to consider path integrals that in a sense zigzag back and forth in time.    For example, given an initial state $\Psi$
and an operator $\O$, one might want to calculate $\la\Psi|\O(\tau)|\Psi\ra=\la \Psi|e^{\i H \tau} \O e^{-\i H \tau}|\Psi\ra$.   To describe this as a path integral,
we need a path integral that propagates the state forwards in time by a time $\tau$ to construct the factor $e^{-\i H\tau}$, after which we insert the operator $\O$ and then apply a path integral
that will propagate the state backwards in time to construct the factor $e^{\i H\tau}$.

As explained in section \ref{allowable}, from the point of view of allowable complex metrics, the two cases of forwards and backwards propagation in time
correspond to two possible regularizations of a Lorentz signature metric, in which, for example, $-\d t^2+\d \vec x^2$ is replaced
by $-(1\mp\i\epsilon)\d t^2 +\d \vec x^2$.   

\begin{figure}
 \begin{center}
   \includegraphics[width=2.1in]{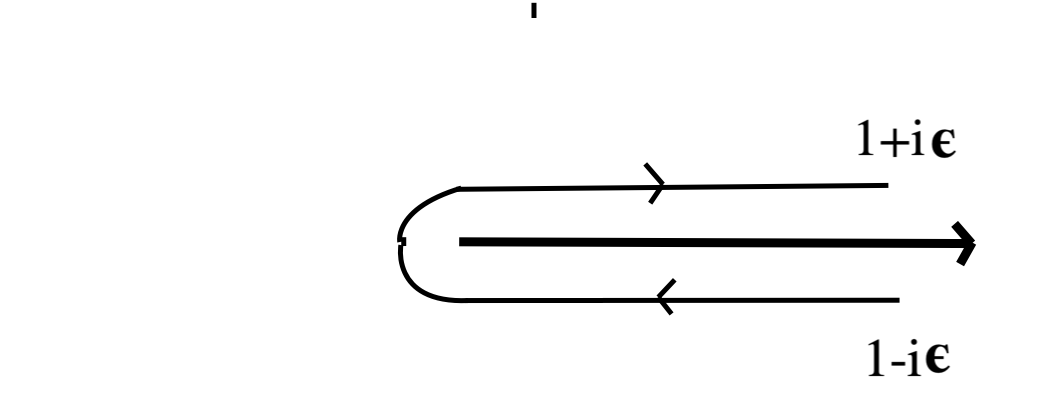}
 \end{center}
\caption{\small The long way around from $1-\i\epsilon$ to $1+\i\epsilon$, avoiding the positive real axis.  \label{Patience}}
\end{figure}

Thus, if we do not mind working with a discontinuous metric, we can describe the timefold with the discontinuous metric
\be\label{telmigo}\d s^2=-f(t) \d t^2+\d \vec x^2,\ee
with
\be\label{elmigo}f(t)=\begin{cases} 1-\i\epsilon & t<0\cr   1+\i \epsilon & t>0 .\end{cases}\ee
However, to couple quantum fields to a discontinuous metric is likely to be problematical.
To get something better behaved, we should replace $f(t)$ with a smooth function that equals
$1-\i\epsilon$ for (say) $t<-\delta$, for some small $\delta$, and $1+\i\epsilon$ for $t>\delta$.   The only subtlety is that $f(t)$ is not
allowed to pass through the positive real axis, as this would violate the condition (\ref{mineq}) for an allowable metric.  
So $f(t)$ has to go ``the long way around'' from $1-\i\epsilon$ in the lower half plane through the negative real axis and finally to $1+\i\epsilon$ in the upper
half plane (fig. \ref{Patience}).   The portion of this path in which $\Re\,f(t)<0$ gives some imaginary time propagation that 
regularizes the real time path integral that we would
naively compute using the real but discontinuous function $f(t)$ in eqn. (\ref{elmigo}).

\subsection{The Double Cone}\label{doublecone}

Somewhat similar to a timefold is the double cone, which has been used  \cite{SSS}
to study the spectral form factor $ \la \Tr\,e^{\i H T} \,\Tr \,e^{-\i HT}\ra$ in a holographic theory.
A simple version of the double cone metric in two dimensions is
\be\label{dc} \d s^2=-\sinh^2r\, \d t^2+ \d r^2. \ee
If  $t$ is a real variable with an identification $t\cong t+T$, and $r$ is real and nonnegative, then this metric describes a cone, with Lorentz signature and with
a singularity at $r=0$ (where one identifies points with different values of $t$).  
However, in the application to $\la \Tr\,e^{\i H T} \,\Tr \,e^{-\i HT}\ra$, one wishes to allow negative as well as positive values of $r$.
In this case, still identifying $t\cong t+T$, one has a pair of cones, meeting at their common apex.   In a holographic interpretation, the conformal boundaries of the two
cones are related to the two traces in the spectral form factor.

This rather singular spacetime has a rather simple regularization, essentially discussed in \cite{SSS},
 in which we avoid letting $r$ pass through the origin in the complex plane.
We simply take, for example, 
\be\label{rego} r=u-\i\epsilon,~~u\in \R,\ee
leading to the metric
\be\label{cc}\d s^2=-\sinh^2(u-\i \epsilon) \d t^2 + \d u^2. \ee
Here $\epsilon$ is a small nonzero real number, whose sign is not important, in the sense that one can compensate for changing the sign of $\epsilon$
by $u\to -u$, $r\to -r$.   As will be clear in a moment, this would have the effect of reversing which of the two cones computes $\Tr\,e^{-\i HT}$ in a
holographic context and which computes $\Tr\,e^{\i HT}$.

The function $\sinh (u-\i \epsilon)=\cos\epsilon \sinh u-\i\sin \epsilon \cosh u$ is never positive for real $u$, so $-\sinh^2(u-\i \epsilon)$ is never negative, and
 therefore  the line element in eqn. (\ref{cc}) corresponds
to an allowable complex metric.  For positive $u$, $\sqrt{\det g}$ is close to the positive imaginary axis, and for negative $u$, it is close to the negative imaginary axis.  
Accordingly, with standard conventions, the positive $u$ cone is related in a holographic description to $\Tr\,e^{-\i HT}$, and the negative $u$ cone is related
to $\Tr\,e^{\i HT}$.    

Also analyzed in \cite{SSS} is a regularized version of the double cone that is related to $\la \Tr\,e^{-H(\beta-\i T)}\,\Tr\,e^{-H(\beta+\i T)}\ra$.   
The line element is
\be\label{regcone} \d s^2=-\left(\sinh r+\frac{\i\beta}{T}\cosh r\right)^2 \d t^2 +\d r^2,\ee
Again this is an allowable complex metric.

\subsection{Rotating Black Holes}\label{rotating}

The last example that we will consider is a rotating black hole.

Gibbons and Hawking \cite{GH} computed the thermodynamic properties of a Schwarzschild black hole by computing the action of a smooth
Euclidean signature solution of Einstein's equations obtained by continuing the Schwarzschild solution to imaginary time.   They then considered a rotating or Kerr black
hole, and observed that in this case
 continuation to imaginary time gives a complex metric,\footnote{This statement assumes that one continues to imaginary time while keeping the angular momentum
 real.  If one also takes the angular momentum to be imaginary, then there is a real solution in Euclidean signature.   One approach to black hole thermodynamics
 is to compute for imaginary angular momentum using a real Euclidean metric
 and then continue back to real angular momentum.   This seems to give sensible results.   However,
 it is natural to ask what happens if we keep the angular momentum real.  The physical meaning of the imaginary angular momentum ``ensemble''
 based on $e^{-\beta(H-\i |\Omega|J)}$ is not very transparent. We put the word ``ensemble'' in quotes because 
 the operator  $ e^{-\beta(H-\i |\Omega|J)}$ is not positive, so in taking its trace, we are not counting states with positive weights.  It is not clear what states
 dominate this trace, or even whether this question has a clear answer.}
  which they called quasi-Euclidean.    It turned out that the thermodynamic properties
of a Kerr black hole can be computed from the action of the quasi-Euclidean solution, with physically sensible results that are in accord with other approaches.

From the perspective of the present article, it is natural to ask if the quasi-Euclidean metric is an allowable complex metric.   In general, the answer to this
question is ``no.''   For example, let us consider the case of a black hole in an asymptotically flat spacetime.    In the field of a stationary, rotating black hole,
a quantum field has at least two conserved quantities -- the energy $H$ and angular momentum $J$.   (Above $3+1$ dimensions, there might be more than
one conserved angular momentum component.)   The partition function of a quantum field propagating in the black hole spacetime is a contribution
to $\Tr\,\exp(-\beta(H-\Omega J))$, where $\beta $ and $\Omega$ are the inverse temperature and the angular velocity of the black hole.   
In the case of a black hole in asymptotically flat spacetime, a particle of given energy can have arbitrarily large $J$ if it is located far from the black hole.
Hence such a particle can make an arbitrarily large contribution to $\Tr\,\exp(-\beta(H-\Omega J))$, and one should not expect to get a sensible answer
for a quantum contribution to $\Tr\,\exp(-\beta(H-\Omega J))$.   Hence it is natural that the coupling of quantum fields to the quasi-Euclidean metric would be ill-defined,
and that this metric would not be allowable.

In Minkowski space, for example, if $J$ corresponds to the rotation generator $x\partial_y-y\partial x$, then $H-\Omega J$ corresponds to the vector field
\be\label{tellme}V= \frac{\partial}{\partial t}-\Omega\left( x \frac{\partial}{\partial y}-y\frac{\partial}{\partial x}\right). \ee
This vector field is spacelike for $\Omega^2(x^2+y^2)>1$, and hence a particle localized at $\Omega^2(x^2+y^2)>1$ can have an arbitrarily negative
value of $H-\Omega J$.   The same is true in an asymptotically flat  Kerr spacetime.

It was observed in the early days of the AdS/CFT correspondence that matters are better in an asymptotically Anti de Sitter (AAdS) spacetime \cite{HReall, H2,HST}.   In
conformal field theory on a unit sphere, the operator $H-\Omega J$ is bounded below if $|\Omega|<1$, and the trace $\Tr\,\exp(-\beta(H-\Omega J))$  converges under
that restriction on $\Omega$. Therefore, in the bulk dual to a boundary conformal field theory, one expects
the partition function $\Tr\,\exp(-\beta(H-\Omega J))$ to make sense for $|\Omega|$ small enough.

In AAdS space,  pick a rotating black hole solution
with the property that 
outside the black hole horizon, the Killing vector field $V$ that corresponds to $H-\Omega J$ is everywhere timelike.  
One expects that to be the condition  that makes the quantum operator $H-\Omega J$ bounded below for perturbations outside the horizon,
ensuring that the trace 
$\Tr\,\exp(-\beta(H-\Omega J))$, taken over quantum fluctuations outside the horizon, is well-defined.  
The path integral on the quasi-Euclidean metric is supposed to be a way to compute that trace.   So under these conditions,
we may hope that
 the quasi-Euclidean metric might be allowable.    In fact, as we will explain,  the quasi-Euclidean metric is allowable if and only if the vector field $V$ is
 everywhere timelike outside the horizon.   

In four dimensions, a rotating black hole is conveniently parametrized by coordinates $t,\phi,r,\theta$, where $H$ and $J$ are generated by $\partial_t|_{\phi,r,\theta}$
and $\partial_\phi|_{t,r,\theta}$, respectively ($\phi$ is an angular variable with $\phi\cong \phi+2\pi$), and in the asymptotic region, $r$ is a radial coordinate and $\theta,\phi$ are polar angles.   In three dimensions, one can
omit $\theta$ from this discussion; above four dimensions, some additional coordinates are needed but the essence of the following argument is not changed,
as we will explain at the end.  A general form of the metric of a rotating black hole is 
\be\label{genform}\d s^2 = - N^2 \d t^2 +\rho^2 (N^\phi \d t+\d \phi)^2 +g_{rr}\d r^2 +g_{\theta\theta}\d\theta^2.\ee
All functions $N, N^\phi,\rho, g_{rr}, $ and $g_{\theta\theta}$ depend on $r,\theta$ only. This form is partly constrained by the fact that a rotating black
hole has a symmetry under $t,\phi\to -t,-\phi$. We have assumed a coordinate choice such that  $g_{r\theta}=0$, though this will not be important.
The function $N^\phi$ vanishes at $r=\infty$; this is part of the AAdS condition.  On the horizon, $N^\phi$ has a constant value $N^\phi_h$; this fact is
important in constructing the quasi-Euclidean spacetime.  Indeed, the constant $N^\phi_h$ is the quantity $\Omega$ that appears in the black hole
thermodynamics.   It is convenient in constructing the quasi-Euclidean spacetime to define a new angular coordinate $\tilde\phi=\phi+\Omega t$.   The line element is then
\be\label{genuform}\d s^2 = - N^2 \d t^2 +\rho^2 ((N^\phi -\Omega) \d t+\d \tilde \phi)^2 +g_{rr}\d r^2 +g_{\theta\theta}\d\theta^2.\ee
In this coordinate system, the vector field that generates $H-\Omega J$ is just $V=\partial_t|_{\tilde\phi,r,\theta}$.   The condition for $V$ to be
everywhere timelike outside the horizon is therefore simply  $g_{tt}<0$ or
\be\label{enform} N^2-\rho^2(N^\phi-\Omega)^2>0. \ee
This function vanishes on the black hole horizon, where $V$ becomes null.

 In the coordinate system $t,\tilde\phi,r,\theta$, the quasi-Euclidean
metric is constructed by discarding the region behind the horizon, setting $t=\i \tau$, and identifying points on the horizon that differ only in the value
of $\tau$.   The quasi-Euclidean metric is thus 
\be\label{xenform}\d s^2 =  N^2 \d \tau^2 +\rho^2 (\i(N^\phi -\Omega) \d \tau+\d \tilde \phi)^2 +g_{rr}\d r^2 +g_{\theta\theta}\d\theta^2.\ee
In showing that this is smooth, one uses the vanishing of $N^\phi-\Omega$ on the horizon.

It is rather immediate from the definitions that a metric of this form is allowable if and only if, treating $r$ and $\theta$ as constants,
the purely two-dimensional metric $g_{(2)}$ that corresponds to the line element
\be\label{lenform} \d s_{(2)}^2 =  N^2 \d \tau^2 +\rho^2 (\i(N^\phi -\Omega) \d \tau+\d \tilde \phi)^2 \ee
is allowable.   This condition is trivial on the horizon, where $g_{(2)}$ is Euclidean.   Away from the horizon, we have a line element of the general form
\be\label{zenform}\d s_{(2)}^2=A\d \tau^2+B\d\tilde\phi^2+2\i C\d\tau \d\tilde\phi,\ee
with real $A,B,C$.  Such a metric is allowable if and only if $A,B>0$. To see necessity, observe that such a metric has $\det{g_{(2)}}>0$, so the necessary condition
$\Re\,(g_{(2)}/\sqrt{\det g_{(2)}})>0$ for allowability reduces to $\Re\,g_{(2)}>0$, that is, $A,B>0$.  Sufficiency of $A,B>0$ can be seen by observing that  for $A,B>0$,
$g_{(2)}$ can be put in the form
(\ref{diagbasis}) with $|\Arg\,\lambda_1|=|\Arg\,\lambda_2|<\pi/2$.

  The condition $A>0$ is equivalent to eqn. (\ref{enform}), and it is also true that 
$B>0$ outside the horizon, since the black hole metric would fail to have Lorentz signature if $B=\rho^2$ is not positive outside the horizon.    Thus we have shown that the
quasi-Euclidean metric is allowable if and only if the vector field $V$ that generates $H-\Omega J$ is everywhere timelike outside the horizon.

Above four dimensions, rotating black holes are more complicated \cite{MP}.   However, the complications do not really affect the preceding analysis.
Consider a black hole solution that has a Killing vector field that is everywhere timelike outside the horizon, and pick coordinates so that this Killing
vector field is just $\partial/\partial t$; denote the other coordinates as $x^1,x^2,\dots, x^{D-1}$.   The metric then has the general form
\be\label{genfo}\d s^2=g_{tt}\d t^2+\d t\sum_i \alpha_i \d x^i+ h_{ij}\d x^i \d x^j, \ee
where $g_{tt}$, $\alpha_i$, and $h_{ij}$ depend only on $x^1,x^2,\cdots, x^{D-1}$.  The function $g_{tt}$ vanishes on the horizon and is negative outside, and $h_{ij}$ is
positive-definite outside the horizon.    The quasi-Euclidean solution is obtained as before by substituting $t\to \i\tau$, omitting the region behind the horizon, and identifying points on the horizon that differ only in the value of
$\tau$.   Alowability is a pointwise criterion, and in checking this criterion  at a given point, only one linear combination of the $\d x^i$ is relevant,
namely $\sum_i \alpha_i \d x^i$.   Therefore, allowability of the quasi-Euclidean metric again comes down to the fact that a two-dimensional metric of
the form  (\ref{zenform}) is allowable for $A,B>0$.

A simple example of a rotating black hole with a Killing vector field that is everywhere timelike outside the horizon is  the BTZ black hole in three dimensions.
Consider a BTZ black hole of mass $M$ and angular momentum $J$ in a world of radius of curvature $l$.  Black hole solutions exist for $|J|<M l$, with
\begin{align}\label{stuff} N^2=& \left(\frac{r }{l\rho}\right)^2 (r^2-r_+^2) \cr
                                       N^\phi=&-\frac{4GJ}{\rho^2} \cr
                                           \rho^2=&r^2+4GM l^2-\frac{1}{2} r_+^2 \cr
                                              r_+^2=&8Gl\sqrt{M^2l^2-J^2} . \end{align}  The horizon is at $r=r_+$, so $\Omega=-8GJ/(r_+^2+8GM l^2)$.
A short calculation reveals that
\be\label{shortone} N^2-\rho^2(N^\phi-\Omega)^2= \frac{(r^2-r_+^2)}{\rho^2}\left( \frac{r^2}{l^2}+\frac{(4GJ)^2(r^2-r_+^2)}{(\frac{1}{2}r_+^2+4GM l)^2}\right), \ee
which is positive outside the horizon, that is for $r>r_+$.

\section{Searching for the Integration Cycle of the Gravitational Path Integral}\label{cycle}

In \cite{GHP}, Gibbons, Hawking, and Perry (GHP) observed that the Einstein action in Euclidean signature is unbounded below.
In fact, if one makes a Weyl transformation of the metric by $g\to e^{2\phi}g$, the action picks up a negative term proportional
to $\int_M \d^Dx \sqrt{\det g} \,e^{(D-2)\phi} g^{ab}\partial_a\phi \partial_b \phi$.   To deal with this, GHP proposed to Wick rotate the integration contour for
the scale factor of the metric tensor, setting $\phi=\i\varphi$ with $\varphi$ real.    They considered the path integral for asymptotically flat metrics
on a space that is asymptotic to $\R^D$ at infinity, and argued that every such metric can be uniquely written\footnote{In situations other
than asymptotically flat metrics on $\R^D$, one needs to somewhat modify the GHP proposal.   Some
issues concerning this proposal were discussed in \cite{HalHar}.}
as $g=e^{2\phi} g_0$, where $g_0$ is a metric of zero scalar curvature.
GHP formulated a ``positive action conjecture,'' according to which the Einstein action is nonnegative for an asymptotically flat metric $g_0$
of zero scalar curvature.   The combination of the positive
action conjecture for $g_0$ and the contour rotation for $\phi$ was supposed to make the gravitational path integral convergent.  The positive action 
conjecture was later proved by Schoen and Yau \cite{SY}.

At least in the context of perturbation theory around a classical solution, the GHP procedure does make sense of the gravitational path integral,
modulo the usual problems concerning ultraviolet divergences.    In the Gaussian approximation, setting $\phi=\i\varphi$, 
with real $\varphi$, makes the action positive and the path integral
convergent.    In perturbation theory, one is always integrating the 
product of a polynomial times a Gaussian function, and,
provided that the Gaussian is convergent, such an integral is well-defined.  So there are no further difficulties in perturbation theory, 
except for the usual ultraviolet divergences of
quantum gravity. 

Should one do better?
One possible point of view is that the gravitational path integral only makes sense in perturbation theory around a classical solution, and to do better
requires a better theory.   
However, it is also imaginable that extending the GHP recipe to make sense beyond perturbation theory would be a step towards a better theory.
To go beyond perturbation theory, one would want an integration cycle $\Gamma$ in the space of complex-valued metrics such that the real part of the Einstein action grows
at infinity along $\Gamma$ -- ensuring at least formally that  the gravitational path integral converges as an integral on $\Gamma$.   
Extrapolating from \cite{GHP}, an obvious guess  might   be 
to define $\Gamma$ by saying $g= e^{2\i\varphi}g_0$ where $\varphi$ is an angle-valued field and $g_0$ is a real metric of zero scalar curvature.   This is not satisfactory
because with this choice,  the gravitational action  $\int \d^Dx \sqrt g R$  has no useful positivity property; it changes by an arbitrary phase when $\varphi$
is shifted by a constant.

From the perspective of the present paper, one would like  $\Gamma$  
to be contained 
within the space of allowable complex metrics.   
One can  attempt to use gradient flow to construct the integration cycle $\Gamma$,
as described in detail for three-dimensional Chern-Simons  theory in \cite{AnalyticContinuation}.    In perturbation theory, this procedure will be equivalent to that
of \cite{GHP}, but it might extend beyond perturbation theory.
The general procedure is as follows.  Let $\Phi^I $, $I=1,\cdots,N$ be a set of fields or integration
variables, with an action $I(\Phi^I)$, so that the integral of interest is formally $Z_U=\int_U \d\Phi^1\cdots\d\Phi^N \exp(-I(\Phi))$.    The goal is
to generalize this integral, which may not converge,  to a convergent complex contour integral.  As a first step, analytically continue the $\Phi^I$ to complex variables $\varPhi^I$, and analytically continue
the action $I(\Phi) $ to a holomorphic function $\I(\varPhi)$. The $\Phi^I$ are functions on a space $U$ and the $\varPhi^I$ are holomorphic functions on
a complexification $\U$ of $U$.   The goal is now to find a middle-dimensional integration cycle $\Gamma\subset\U$, such that the integral
$Z_\Gamma=\int_\Gamma\d\varPhi^1\cdots\d\varPhi^N\exp(-I(\varPhi))$, which formally reduces to the original $Z_U$ if $\Gamma=U$, converges.

This may formally be done as follows.\footnote{See \cite{TL} for a discussion of this formalism in the context of gravity (and an introduction to the formalism). The perspective taken there
is that the Lorentz signature path integral is the basic definition, and analytic continuation to complex metrics is used only as a procedure to more
precisely define and evaluate it and deal with any ambiguities.  A consequence is that complex critical points can only make exponentially small
contributions, not exponentially large ones.   So in particular the $e^{S/2}$ enhancement in the ``creation of a universe from nothing,'' discussed in section
\ref{hhw}, is replaced by an $e^{-S/2}$ suppression.}
  Pick a positive-definite metric $G$ on $\U$.     In the previously mentioned application of
this formalism to Chern-Simons theory,
 a simple choice of $G$ leads to a relation with renormalizable gauge theory in four dimensions.  In gravity, since we are dealing anyway with a highly nonlinear
 low energy effective
 field theory, we can contemplate a rather general choice of $G$. We will, however, assume that $G$ is a  Kahler metric $G_{I\bar J}\d \varPhi^I\d\bar\varPhi{}^{\bar J}$, 
 as this leads to some simplifications.  Introduce a ``flow variable,'' a real variable $s$, and view the $\varPhi^I$ as 
 functions of $s$. 
 Now consider the gradient flow equation
 \be\label{gradflow}\frac{\d \varPhi^I}{\d s}=G^{I\bar J}\frac{\partial\Re\, \I}{\partial \bar\Phi^{\bar J}}.\ee
 In an application to $D$-dimensional field theory on a $D$-manifold $M$, this equation is really a differential equation on a $D+1$-manifold $M\times \R$, where
 $\R$ is parametrized by $s$.   The solutions of the gradient flow equation in which  $\varPhi$ is independent of $s$ correspond to critical points of the action function $\I$ or in other words to
 classical solutions of the equations of motion.   In any nonconstant solution, $\Re\,\I$ is a strictly increasing function of $s$.  
 
 Let $p$ be a critical point of the action,\footnote{In all of the following statements, one has to allow critical points at infinity as well as ordinary critical points.
 Critical points at
 infinity were discussed in section \ref{hhw}.}
 which for simplicity we will assume  to be isolated and nondegenerate. (The discussion can be generalized to include other cases.)   Because the function $\Re\,\I$ is the real part of a holomorphic function, the matrix of second derivatives of $\Re\,\I$ at $p$, which is
 known as the Hessian matrix,  has equally many 
 positive and negative eigenvalues.\footnote{For example, if $\U$ has complex dimension 1, we can pick a local holomorphic coordinate $z=x+\i y$ on $\U$ 
 near a critical point so that $\I=z^2$.   Then $\Re\,\I=x^2-y^2$, and clearly the Hessian matrix of  this function has one positive and one
 negative eigenvalue.}  Now consider solutions of the gradient flow equation on a semi-infinite interval $-\infty<s\leq 0$ that start at $p$ at $s=-\infty$.   
 Because half the eigenvalues of the Hessian at $p$ are positive and half are negative, the  values of such a solution at $s=0$ comprise a middle-dimensional submanifold $\Gamma_p\subset \U$.  In a finite-dimensional context, under mild
 assumptions,\footnote{The main assumption needed is that  there is no solution of the flow
 equation on the whole real line $-\infty<s<\infty$ that flows from one critical point $p$ at $s=-\infty$ to another one $p'$ at $s=+\infty$.   A sufficient condition 
 to ensure that no such flow exists
  is that the different critical points have different values of ${\mathrm{Im}}\,\I$. This statement depends on the metric $G$ being Kahler.
   If there is a flow between two distinct critical points $p$ and $p'$, one says that
 one is sitting on a Stokes line (in the space of all possible actions $\I$), and the statements in the text require some modification.  }
 the function $\Re\,I$ grows at infinity along $\Gamma_p$ and the integral $\int_{\Gamma_p}\D\varPhi^1\cdots \D\varPhi^N \exp(-\I(\varPhi))$ converges.
 Moreover, again under some mild assumptions, any integration cycle $\Gamma$ such that the integral $\int_{\Gamma}\D\varPhi^1\cdots \D\varPhi^N \exp(-\I(\varPhi))$
 converges is an integer linear combination of the $\Gamma_p$.  
 
 In gravity, $U$ would be the space of all metrics on a manifold $M$ and $\U$ is the space of complex-valued metrics on $M$.   The diffeomorphism group $\Diff(M)$ 
 of $M$ acts on $U$, and the path integral is really an integral over $U/\Diff(M)$. Upon complexification, $U$ is replaced by $\U$.  One might naively think
 that $\Diff(M)$ would have a complexification $\Diff_\C(M)$ and that one would really want to define a cycle in $\U/\Diff_\C(M)$.   This is wrong for two
 reasons.  First, although the Lie algebra $\diff(M)$ of $\Diff(M)$ can be complexified to a complex Lie algebra $\diff_\C(M)$, 
 there is no corresponding complexification of the group $\Diff_\C(M)$, so there is no way
 to define a quotient $\U/\Diff_\C(M)$.   Second, even in gauge theory, where a complexification of the gauge group does exist, the gradient flow equation is not invariant
 under this complexification.  The appropriate procedure to deal with gauge symmetries was described in \cite{AnalyticContinuation}.  One replaces the action $\I(\varPhi)$
 with an extended action $\I(\varPhi)+\int_M\phi\mu$, where $\phi$ is a Lagrange multiplier and $\mu$ is a moment map for the imaginary part of $\diff_\C(M)$.
 Roughly, $\mu=0$ is a partial gauge-fixing condition that reduces invariance under $\diff_\C(M)$ to invariance under $\diff(M)$.   Then one proceeds as before,
 studying the flow equation of this extended action.
 
 Unfortunately, it seems doubtful that this procedure will really accomplish what we want in the case of gravity.
 With a simple choice of the metric $G$, there is no obvious reason for $\Gamma$ to remain in the space of allowable metrics.
 We could choose $G$ to be a complete Kahler metric on the space of allowable metrics.   This will force $\Gamma$ to remain in the space of allowable
 metrics, but the real part of the action might not go to $+\infty$ along $\Gamma$.

 An alternative to this discussion -- and to the framework of the present article -- would be to use gradient descent to define an integration cycle not for 
 gravity alone but for the combined system of gravity plus matter fields.   One would simply follow the above-described procedure, but taking $\I$ to be the combined
 action  of gravity plus matter.   In this case, all critical points are potentially allowed; one simply Wick rotates all gravitational or matter
 field variables to make any integral converge.  Another mechanism has to be found to exclude undesirable examples such as those of 
 section \ref{examples}.  As remarked
 in section \ref{allowable}, the sum over discrete fluxes for matter fields may be a problem in such an approach, and one also would have to accept that in expanding
 around a critical point, the matter path integral does not have a Hilbert space interpretation.   The reason for the last statement  is for a path integral on a manifold $M$
 to have a Hilbert space interpretation, the field variables and integration cycle have to be defined by local conditions on $M$; an integration cycle produced by
 gradient descent (or in perturbation theory
 by Wick rotating any mode whose kinetic energy has a real part with the wrong sign) does not have the appropriate locality.  Still another issue is that Wick rotating the matter fields multiplies the path
integral measure by a potentially ill-defined phase. 

One last comment concerns the positive action theorem.  Consider instanton solutions of Einstein's equations with zero cosmological constant
that are asymptotic to $\R^D$ at spatial infinity.
 It is not difficult to prove that such a solution has zero action.\footnote{The bulk term $\int \d^D x \sqrt g R$
in the action vanishes for a Ricci flat metric, and  the linearized Einstein's equations imply that a solution that is asymptotic to flat $\R^D$
at spatial infinity approaches the flat metric on $\R^D$ fast enough that the Gibbons-Hawking-York surface term in the action also vanishes.}    It would presumably
not be physically sensible to include in the gravitational path integral an instanton with zero action, since its contribution to the path integral would be too large, so one is led to hope that such solutions do not exist.
Indeed, a special case of the positive action conjecture of \cite{GHP}, proved in \cite{SY}, says that the Einstein equations in Euclidean signature with zero cosmological
constant
have no solution asymptotic to $\R^D$ other than $\R^D$ itself.   With complex metrics, this is not so, as we saw in an example in section \ref{examples}.
Optimistically,  an extension of the positive action theorem might say that the real part of the action is always positive for an allowable classical
solution that is asymptotic to $\R^D$ at infinity.  Alternatively, perhaps  allowability is not the right condition or not a sufficient condition.

\noindent{\it Acknowledgments}  I thank R. Bousso,  J.-L. Lehners, L. Iliesiu,
 J. Maldacena, S. Murthy, H. Reall,  G. Segal, S. Shenker, R. Sorkin,  M. Taylor, and J. Turiaci  for helpful discussions. Research supported in part by
 NSF Grant PHY-1911298.

\appendix

\section{Pontryagin Classes, Euler Classes, and Volumes}\label{gb}

Here we discuss topological aspects of the strange world of nondegenerate but possibly nonallowable complex metrics.

The Chern classes of a rank $N$ 
complex vector bundle $V$ over a manifold $X$ can be described in de Rham cohomology by familiar expressions involving the curvature $F$ 
of any connection $A$ on $V$.   For example, the second Chern class is associated to the four-form $\Tr\,F\wedge F/8\pi^2$.   Depending on the dimension of $X$,
Chern numbers of $V$ can be defined as integrals over $X$ of products of Chern classes.   

This construction is probably most familiar in the case that the structure group of $A$ is $U(N)$, but it is valid more generally if $A$ is a connection with
structure group $GL(N,\C)$.  

If instead $V$ is a rank $N$ real vector bundle over $X$, then its Pontryagin classes are defined as the Chern classes of the complexification $V_\C$ of $X$.
The Pontryagin classes of a manifold $X$ are defined as the Pontryagin classes of its tangent bundle $TX$.   So in other words, they are by definition the
Chern classes of the complexification $T_\C X=TX\otimes _\R\C$ of $TX$, and they can be computed using any connection, in general of structure group
$GL(N,\C)$, on $T_\C X$.   So given any connection $A$ on $T_\C X$ with curvature $F$, the Pontryagin classes of $X$ can be defined by the usual polynomials
in $F$.

In particular,  suppose we are given a nondegenerate complex metric $g$ on $X$, not necessarily allowable.
   The standard formulas defining the Levi-Civita connection on the tangent bundle
of $X$ make sense in this situation, but if $g$ is not real, they give a complex-valued connection.   In other words, the Levi-Civita connection associated to a complex metric
is a connection on $T_\C X$.  The structure group of this connection is in general $O(N,\C)$, a subgroup of $GL(N,\C)$.  But as we have just noted, Pontryagin classes  can be defined using the curvature of any
connection on $T_\C X$.   So in particular, Pontryagin numbers of $X$ can be computed using the Riemann curvature tensor $R$ of any complex-valued metric.
For example, if $X$ is a four-manifold, the integral $-\int_X \Tr\,R\wedge R/8\pi^2$ associated to the first Pontryagin class will always have its standard value.  

One way to understand this statement is the following.   If $g$ is a non-allowable complex metric on $X$, it may not be possible to interpolate continuously
 in the space of invertible complex metrics
between $g$ and a Euclidean metric $g_0$.   But it is always possible to continuously interpolate between the
Levi-Civita connection $\omega$ derived from $g$ and the Levi-Civita connection $\omega_0$ derived from $g_0$:  one just considers the family
of connections on $T_\C X$ defined by $(1-t)\omega+t \omega_0$, $0\leq t\leq 1$.  (The interpolating connections in general have structure group
$GL(N,\C)$.)  So they give the same results for Pontryagin classes.

In dimension $N=2k$, another interesting curvature integral is the integral that appears in the Gauss-Bonnet formula for the Euler characteristic of $X$.
As was noted by Louko and Sorkin\footnote{They illustrated the point with a nondegenerate complex metric on $S^2$ whose curvature vanishes,
so the usual Gauss-Bonnet integral gives the value 0, not the standard Euler characteristic 2 of $S^2$.  Their example of the complex flat metric on $S^2$
was described in section \ref{examples}.}  \cite{SL}
 in the case $k=1$, it is not true that in general the Gauss-Bonnet integral computed using the curvature of a nondegenerate
complex metric takes its standard value.   In the case of the curvature formula for the Euler characteristic, the underlying invariant is the Euler class of an oriented real
vector bundle.   The differential form that represents the Euler class of a real vector bundle is a multiple of $\epsilon^{a_1a_2\cdots a_{2k}} F_{a_1a_2} 
F_{a_3a_4}\cdots F_{a_{2k-1}a_{2k}}$, where we view $F$ as a two-form valued in antisymmetric $2k\times 2k$ matrices (generators of the Lie algebra of $SO(2k)$).
Thus the definition uses the fact that there is a symmetric $k^{th}$ order invariant on the Lie algebra of $SO(2k)$, namely the tensor
$\epsilon^{a_1a_2\cdots a_{2k}}$ (in a different language, the invariant is the Pfaffian of an antisymmetric matrix).
   There is no such invariant on the Lie algebra of
$GL(2k,\C)$ (or even $SL(2k,\C)$).   But there is such an invariant in $SO(2k,\C)$.   So without more structure, a rank $2k$ complex vector bundle $V\to X$
does not have an Euler class.  However, if we are are given a reduction of the structure group of $V$ to $SO(2k,\C)$, then one can define a Euler
class of $V$.   It is not a topological invariant of $V$ but depends on the topological choice of the reduction of structure group to $SO(2k,\C)$.   

Applying this to complex metrics on $X$, we observe that a complex nondegenerate metric determines a reduction of the structure group of $T_\C X$
to $O(2k,\C)$.   If the structure group can be further reduced to  $SO(2k,\C)$ (which is the case if the Levi-Civita connection  derived from $g$ has holonomy
in $SO(2k,\C)\subset O(2k,\C)$), then this enables 
one to define an Euler class of $T_\C X$.  This Euler class can be represented by the standard 
curvature polynomial.   However, the Euler class of $T_\C X$ defined this way in general really will depend on the topological class of the chosen
complex metric, and will not agree with the standard Euler class.

A simple case in which $X$ is orientable but a complex metric $g$ on $X$ reduces the structure group of $T_\C X$ to $O(2k,\C)$ but not to $SO(2k,\C)$ is
as follows.   Let $X$ be a two-torus with angular coordinates $\alpha,\beta$ ($0\leq \alpha,\beta\leq 2\pi$) and consider the metric
$\d s^2=\d \alpha^2+ e^{\i \beta }\d\beta^2$.    A simple computation shows that this metric is flat, and that the holonomy under $\beta\to \beta+2\pi$ is
${\mathrm{diag}}(1,-1)$, which is valued in $O(2,\C)$ but not in $SO(2,\C)$.    Accordingly, it is not possible to define an Euler class with this metric.
We note as well that $\sqrt{\det g}=e^{\i\beta/2}$ likewise changes sign under $\beta\to\beta+2\pi$, so that with this metric it is not possible to define
the volume of $X$. 

 The general situation concerning the obstruction to defining
$\sqrt {\det g}$ for a complex nondegenerate metric is as follows.   A complex vector bundle such as $T_\C X$ does not have a first Stieffel-Whitney class.
However, a reduction of its structure group to $O(N,\C)$, such as that which is provided by a complex invertible metric $g$, enables one to define a first Stieffel-Whitney
class $w_1^g(T_\C X)$.   If and only if this coincides with the usual $w_1(X)$ (the class that measures the obstruction to the orientability of $X$), 
$\sqrt{\det g}$ can be consistently defined.   In the preceding example with $X$ a two-torus, $w_1(T_\C X)=0$ but $w_1^g(T_\C X)\not=0$, and there
is an obstruction to globally defining $\sqrt {\det g}$.   

One way to see that the obstruction to defining $\sqrt{\det g}$ and the obstruction to defining the Euler class must be the same is the following.
The integrand in the Gauss-Bonnet formula for the Euler characteristic
is $\sqrt{\det g}$ times an invariant polynomial in the Riemann tensor $R$.   $R$ is well-defined for any invertible
complex metric, as is any invariant polynomial in $R$.   So the obstruction to making sense of the Gauss-Bonnet formula is the obstruction to defining $\sqrt{\det g}$.

The Euler class  in dimension 2 can be analyzed in more detail as follows.   For simplicity we  consider the case of an orientable two-manifold $X$.   The tangent bundle
$TX$ of $X$ is in general nontrivial as a real vector bundle; the obstruction to its triviality is the Euler class.  But the complexification $T_\C X$ of $TX$ is trivial
and in particular its first Chern class is 0, $c_1(T_\C X)=0$.
Now suppose we are given a complex invertible metric $g$ on $X$.   This determines two null directions in $T_\C X$ and hence  locally it determines
a decomposition of $T_\C X$
as a direct sum of line bundles.   The exchange of the two line bundles under monodromy by the Levi-Civita connection
 would involve an element of $O(2,\C)$ with determinant $-1$.   So if
the metric $g$ actually leads to a reduction of the structure group to $SO(2,\C)$, which is the case that an Euler characteristic can be defined, the
decomposition of $T_\C X$ as a sum of two null line bundles can be made globally.  (In the example given earlier of the metric $\d s^2=\d\alpha^2+e^{\i \beta}
\d\beta^2$, the two null directions correspond to the null 1-forms $\d\alpha\pm \i e^{\i \beta/2}\d \beta$, which are exchanged under $\beta\to\beta+2\pi$,
so the decomposition cannot be made globally.)  
 Since $c_1(T_\C X)=0$, if $T_\C X$ is decomposed as the direct sum of two line bundles, then the form of the decomposition is $T_\C X=\L\oplus \L^{-1}$, for some $\L$.   The complex line bundle $\L\to X$ is classified topologically
 by its first Chern class, which can be an arbitrary integer $n$.   For any $n$, $\L\oplus \L^{-1}$ is trivial, and it is possible to pick a complex metric on $X$
 such that the decomposition in null directions is $T_\C X=\L\oplus \L^{-1}$.   
 The Levi-Civita connection of $X$  in this situation has structure group $\C^*$ (the complexification of the usual $SO(2)=U(1)$) and concretely it can be viewed
 as a connection on $\L$.
 The usual Gauss-Bonnet integral computes the first Chern class of $\L$.   So, generalizing the observation of Louko and Sorkin, the Gauss-Bonnet
 integral  can have any integer value.

\bibliographystyle{unsrt}

\end{document}